%% file: MainArxiv.tex
\documentclass[notitlepage,11pt,times]{article}
\usepackage[letterpaper, total={7in, 9in}]{geometry}
\pdfoutput=1
\usepackage{amsmath}
\usepackage{amssymb}
\usepackage{graphicx}
\usepackage{algorithm}
\usepackage{algpseudocode}
\usepackage{subfig}
\usepackage[T1]{fontenc}

\DeclareMathOperator*{\argmin}{argmin}
\newcommand{\tensor}{\mathbf}
\newcommand{\tp}{^T}
\newcommand{\vectorize}{\text{vec}}
\newcommand{\EE}[1]{\times 10^{#1}}

\begin{document}
\title{Multidimensional NMR Inversion without Kronecker Products: Multilinear Inversion}
\author{David Medell\'in\footnote{Corresponding author. Email: djmedellin@utexas.edu}, Vivek R. Ravi, Carlos Torres-Verd\'in \\
{Department of Petroleum and Geosystems Engineering}\\
{The University of Texas at Austin, 200 E. Dean Keeton, Mail Stop C0300,}\\
{Austin, TX, 78712-1585, USA} }
\date{\today}
\maketitle
\begin{abstract}
Multidimensional NMR inversion using Kronecker products poses several challenges. First, kernel compression is only possible when the kernel matrices are separable, and in recent years, there has been an increasing interest in NMR sequences with non-separable kernels. Second, in three or more dimensions, the singular value decomposition is not unique; therefore kernel compression is not well-defined for higher dimensions. Without kernel compression, the Kronecker product yields matrices that require large amounts of memory, making the inversion intractable for personal computers. Finally, incorporating arbitrary regularization terms is not possible using the Lawson-Hanson (LH) or the Butler-Reeds-Dawson (BRD) algorithms.
We develop a minimization-based inversion method that circumvents the above problems by using multilinear forms to perform multidimensional NMR inversion without using kernel compression or Kronecker products. The new method is memory efficient, requiring less than 0.1\% of the memory required by the LH or BRD methods. It can also be extended to arbitrary dimensions and adapted to include non-separable kernels, linear constraints, and arbitrary regularization terms. Additionally, it is easy to implement because only a cost function and its first derivative are required to perform the inversion.
\end{abstract} 
\input{Introduction}
\input{ProblemStatement}
\input{Method}

\input{Results}
\input{Conclusions}

\section{Acknowledgments}
\label{sec:Acknowledgments}
We are greatly indebted to Prof. Hugh Daigle for giving us access to his laboratory and NMR system. We also thank Joaquin Amb\'ia-Garrido,  Al\'an D\'avila, Wilberth Herrera-Su\'arez, Shaina A. Kelly, and Adam McMullen for proofreading the manuscript. D. Medell\'in also thanks Rafael A. Longoria and Benjamin Crestel for useful discussions and suggestions. The authors would also like to thank Gary Miscoe and Glen Baum for laboratory technical support. We also thank the anonymous reviewers for their suggestions and their meticulous and thorough revision of the manuscript. Financial support for this work was provided by: Anadarko, Aramco, Baker-Hughes, BHP Billiton, BP,  China Oilfield Services LTD., ConocoPhillips, Det Norske, ENI, ExxonMobil, Hess, Paradigm, Petrobras,  Repsol, DEA, Schlumberger, Shell, Statoil, TOTAL, Wintershall and Woodside Petroleum Limited.

\newpage
\input{AbbrevSymb}

\newpage
\bibliographystyle{abbrv} 
\bibliography{References}
\end{document}

%% file: Introduction.tex
\section{Introduction}
\label{sec:Introduction}

Notwithstanding the progress in acquiring multidimensional NMR data \cite{zhang2013fast}, a main challenge continues to be the problem of multidimensional NMR inversion \cite{sun_global_2005,tan2014comparative,zhang2013evaluation,arns2007multidimensional}. While efficient and fast methods exist for the inversion and regularization of 1D NMR data \cite{butler1981estimating}, 2D NMR inversion remains challenging. The usual way of inverting 2D NMR data is to transform the 2D inversion problem into an equivalent 1D inversion problem using Kronecker products. However, after this conversion, the size of the resulting matrix grows with the square of the number of elements in the initial 2D problem. For instance, if the 2D NMR problem required storing $10^4$ numbers, the equivalent 1D problem would require storing $10^8$ numbers. 

To overcome this high memory requirement, \cite{LalithaSolvingFredholm2002} introduced an elegant method that compresses both the NMR data and the kernels of the 2D NMR inversion problem using singular value decomposition (SVD). As a result of the compression, the resulting 1D NMR inversion problem is small enough to be handled by desktop computers. The 1D problem is then inverted using the Butler-Reeds-Dawson (BRD) algorithm \cite{butler1981estimating}. Despite the method's success in inverting 2D NMR data, kernel compression using SVD can only be applied when the NMR kernels are separable -- a condition not satisfied by some pulse sequences \cite{hurlimann_quantitative_2002}. Moreover, SVD is not unique in higher dimensions \cite{kolda2009tensor}, which makes kernel compression ill-defined. Finally, the BRD algorithm can only incorporate zeroth-order regularization.

We propose performing multidimensional linear NMR inversion without converting it into an equivalent 1D NMR inversion problem by minimizing cost functions using a generalized multidimensional version of steepest descent that uses gradients in tensor form (matrices in 2D). Tensor gradients allow us to generalize inversion methods to higher dimensions, avoiding memory expensive Kronecker products used to convert higher dimensional inversion problems to one dimensional ones. To enforce the nonnegativity constraint, we use the gradient projection method \cite{bertsekas_goldstein_levitin_polyak_1976} as opposed to the LH or BRD methods, which are not directly applicable to gradients in tensor form.

The structure of the paper is as follows: in section \ref{sec:StatementProblem} we introduce the problem of multidimensional NMR inversion with a specific example - a $T_1$-$T_2$ NMR inversion. This section has two purposes: first, to define the terms and notation used throughout the paper, and second, to highlight the disadvantages of using Kronecker products to transform the 2D inversion problem to a 1D one.

In section \ref{sec:Method} we develop the multilinear inversion method using the previously introduced $T_1$-$T_2$ NMR inversion. We show how to incorporate multi-order Tikhonov regularization and how to enforce nonnegativity using the projection gradient method. Additionally, we generalize the multilinear inversion to three dimensions and show how to include non-separable kernels and arbitrary regularization terms. 

Finally, in section \ref{sec:Results}, we document five multidimensional inversion examples:

\begin{enumerate}
	\item A 2D $T_1$-$T_2$ NMR inversion with zeroth-order Tikhonov regularization of a laboratory data set of water-saturated Berea sandstone.
	\item A full 2D $T_1$-$T_2$ NMR inversion (no kernel or data compression) with zeroth-order Tikhonov regularization of a laboratory data set of water-saturated Berea sandstone. 
	\item A 3D $D$-$T_1$-$T_2$ NMR inversion with zeroth-order Tikhonov regularization of a synthetic data set representing a rock sample containing a mixture of water and light oil.
	\item A 3D $D$-$T_1$-$T_2$ NMR inversion with zeroth- and second-order Tikhonov regularization of a laboratory data set of a 2\% solution of NaCl in a mixture of water and heavy water (D$_2$O). 
	\item A 3D $D$-$T_1$-$T_2$ NMR inversion with zeroth-order Tikhonov regularization of a laboratory data set of water-saturated Berea sandstone. 
\end{enumerate}

%% file: ProblemStatement.tex
\section{Statement of the Problem}
\label{sec:StatementProblem}
Consider a 2D NMR experiment such as a $T_1$-$T_2$ measurement performed with a CPMG sequence using inversion recovery. Let $f(\tau_2,\tau_1)$ denote the 2D probability density function of protons with longitudinal and transverse relaxation times $\tau_1$ and $\tau_2$, respectively. The NMR signal $S$ at time $t$ after a polarization time $T_W$ is given by
\begin{equation}
S(t,T_W)=S_0\iint{ K_1(T_W,\tau_1) K_2(t,\tau_2)f(\tau_2,\tau_1) d\tau_1 d\tau_2},
\label{eq:T1T2signal}
\end{equation}
where $S_0$ is the maximum strength of the NMR signal for a completely polarized sample and $K_1(T_W,\tau_1)$ and $K_2(t,\tau_2)$ are the $T_1$ and $T_2$ kernels defined by
\begin{eqnarray}
K_1(T_W,\tau_1) & = & 1-2e^{-T_W/\tau_1}, \\
K_2(t,\tau_2) & = &  e^{-t/\tau_2}.
\label{eq:T1T2Kernels}
\end{eqnarray}

We discretize \eqref{eq:T1T2signal} and convert it to a double sum by writing the kernels and the probability density as
\begin{equation}
S(t_i,T_{Wj})=\sum_k{\sum_q K_{2}(t_i,\tau_{2k}) K_{1}(T_{Wj},\tau_{1q}) F(\tau_{2k},\tau_{1q})},
\label{eq:T1T2signalSums}
\end{equation}
where $F(\tau_{2k},\tau_{1q})=S_0\cdot f(\tau_{2k},\tau_{1q})\Delta\tau_{1q}\Delta\tau_{2k}$. In matrix form, \eqref{eq:T1T2signalSums} can be written as
\begin{equation}
\tensor{S}=\tensor{K_2}\tensor{F}\tensor{K_1\tp},
\label{eq:T1T2signalTensor}
\end{equation}
where the components of the matrices are given by
\begin{eqnarray}
(\tensor{K_1})_{jq} &=& 1-2\exp(-T_{Wj}/\tau_{1q}), \\
\label{eq:TensorKernelT1}
 (\tensor{K_2})_{ik} &=& \exp(-t_i/\tau_{2k}),  \\
\label{eq:TensorKernelT2}
\tensor{F}_{kq} &=& S_0\cdot f(\tau_{2k},\tau_{1q})\Delta\tau_{1q}\Delta\tau_{2k}.
\label{eq:TensorF}
\end{eqnarray}
Given a set of NMR data $\tensor{D}$ and kernels $\tensor{K_1}$ and $\tensor{K_2}$, we seek a matrix $\tensor{F}$ for which $\tensor{S}$ best approximates $\tensor{D}$, i.e., to solve the matrix equation with unknown matrix $\tensor{F}$:
\begin{equation}
\tensor{D}\sim\tensor{K_2}\tensor{F}\tensor{K_1\tp}.
\label{eq:InversionEquation}
\end{equation}
This is typically achieved via minimization of the cost function
\begin{equation}
C(\tensor{F})=\left\| \tensor{K_2}\tensor{F}\tensor{K_1\tp} - \tensor{D} \right\|_F^2 ,
\label{eq:CostFunction}
\end{equation}
 with the constraint that $\tensor{F}$ must be a nonnegative matrix (for all $(k,q)$, $\tensor{F}_{kq}\geq0$). Here, $\left\| \cdot \right\|_F$ denotes the Frobenius norm of a matrix. The inversion of \eqref{eq:InversionEquation} is an ill-posed problem as it arises from a Fredholm integral equation of the first kind; its inversion requires regularization. The most common type of regularization is zeroth-order Tikhonov regularization (although other regularization schemes are possible, \cite{chouzenoux2010efficient}). While the BRD algorithm provides an elegant way of inverting NMR data with nonnegative constraints and zeroth-order Tikhonov regularization, it is only directly applicable to 1D NMR inversion problems.

Consequently, \eqref{eq:CostFunction} is usually written in 1D form by transforming the matrix $\tensor{F}$ into a column vector by lexographically stacking the columns of $\tensor{F}$ into a single column vector, an operation known as vectorization. This produces a column vector $\tensor{F_A}$. Accordingly, matrix $\tensor{D}$ is vectorized to form the column vector $\tensor{D_A}$, while the kernels $\tensor{K_1}$ and $\tensor{K_2}$ are combined using the Kronecker product to produce the augmented matrix $\tensor{K_A}=\tensor{K_1}\otimes\tensor{K_2}$. The triple matrix product in \eqref{eq:CostFunction} takes the form
\begin{equation}
\vectorize(\tensor{K_2}\tensor{F}\tensor{K_1\tp})=(\tensor{K_1}\otimes\tensor{K_2})\vectorize(\tensor{F})=\tensor{K_A}\tensor{F_A},
\label{eq:ModelVector}
\end{equation}
where $\tensor{K_1}$ is a $M_1\times N_1$ matrix ($\tensor{K_1}\in\Re^{M_1\times N_1}$), $\tensor{K_2}$ is a $M_2\times N_2$ matrix ($\tensor{K_2}\in\Re^{M_2\times N_2}$), $\tensor{K_A}$ is a $(M_1\cdot M_2)\times (N_1\cdot N_2)$ matrix ($\tensor{K_A}\in\Re^{(M_1\cdot M_2)\times (N_1\cdot N_2)}$), $\tensor{F_A}$ is a column vector of size $N_1\cdot N_2$ ($\tensor{F_A}\in\Re^{(N_1\cdot N_2)\times 1}$), and  $\tensor{D_A}$ is a column vector of size $M_1\cdot M_2\times 1$  ($\tensor{D_A}\in \Re^{(M_1\cdot M_2)\times 1}$).
The 1D augmented cost function, $C_A(\tensor{F_A})$, is therefore 
\begin{equation}
C_A(\tensor{F_A})=\left\| \tensor{K_A}\tensor{F_A}- \tensor{D_A} \right\|^2.
\label{eq:CostFunctionOneD}
\end{equation}
 For a typical laboratory $T_1$-$T_2$ sequence, one has 40,000 echoes ($M_2=N_{echoes}=4\EE{4}$) and 30 $T_1$ polarization steps ($M_1=N_{T_W}=30$). If we attempt to produce a $T_1$-$T_2$ map with 100 points for both $T_1$ and $T_2$ ($N_1=N_2=100$), the matrix $\tensor{K_A}$ would contain $1.2\EE{10}$ numbers. Storing and manipulating matrices of this size poses serious challenges for most desktop computers, especially if the minimization algorithm requires to repeatedly compute the gradient
\begin{equation}
\nabla(C(\tensor{F_A}))=2\tensor{K_A\tp}(\tensor{K_A}\tensor{F_A}-\tensor{D_A}).
\label{eq:GradientK0}
\end{equation}

Mainstream 2D inversion schemes use SVD to compress the kernel and data matrices so that the resulting augmented matrix can be handled by desktop computers. After compression and conversion to a 1D problem, the LH \cite{lawson1974solving} or Venkataramanan's version of the BRD \cite{LalithaSolvingFredholm2002} method can be applied. Figure \ref{fig:Figure_MethodsComparison}a outlines the mainstream 2D inversion scheme.  

The mainstream scheme has several problems. Kernel compression is only possible when the kernel matrices are separable, and in recent years, there has been an increasing interest in NMR sequences with non-separable kernels. To complicate matters, the SVD is not uniquely defined for three or more dimensions. Without kernel compression, the Kronecker product yields matrices that require large amounts of memory, making the inversion intractable for common computers. Moreover, incorporating arbitrary regularization terms is not possible with the LH or BRD method.

%% file: Method.tex
\section{Method}
\label{sec:Method}
\begin{figure}[!htb]
	\centering
		\includegraphics{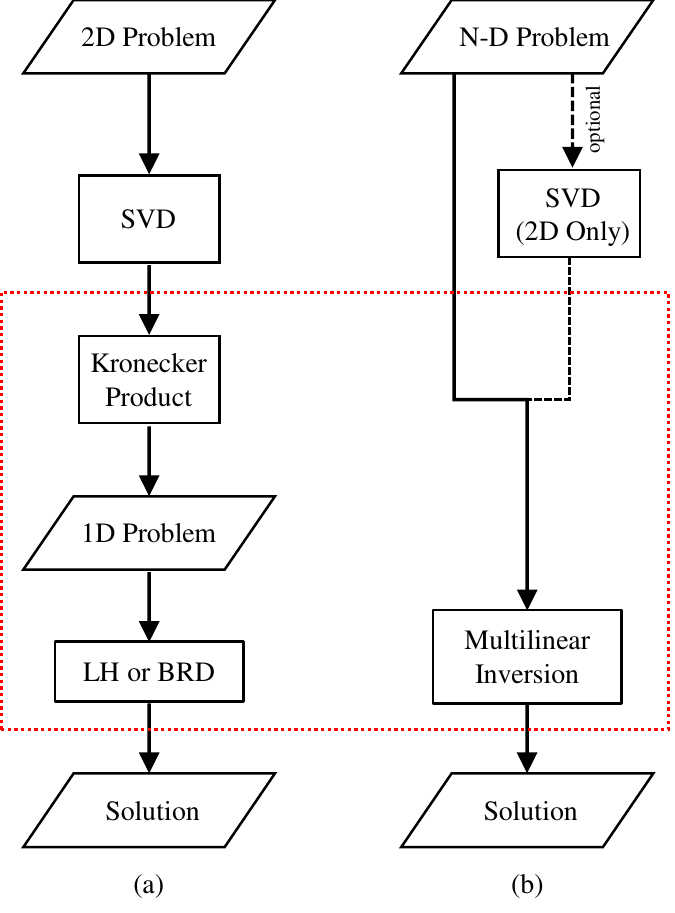}
	\caption[01 Multidimensional linear inversion methods]{Multidimensional linear inversion methods. (a) In the conventional 2D inversion method, the 2D problem is first \textit{compressed} using SVD, then \textit{converted} to a 1D problem using a Kronecker product, and finally inverted using a 1D inversion method such as the LH or the BRD method. (b) Multilinear inversion directly inverts the multidimensional problem without any compression or conversion steps. The dashed box highlights the difference between both methods. Note that SVD compression can also be combined with multidimensional inversion. While in conventional 2D inversion SVD compression is mandatory, in multilinear inversion it is entirely optional.}
	\label{fig:Figure_MethodsComparison}
\end{figure}

Our approach (Figure \ref{fig:Figure_MethodsComparison}b) is to avoid Kronecker products by generalizing the concept of a gradient as a row (or column) vector to a higher dimensional tensor - which we will call \textit{tensor gradient} to distinguish it from the conventional gradient. The tensor gradient enables one to implement a \textit{multilinear solver} (bilinear solver for 2D) using steepest descent to minimize the cost function \eqref{eq:CostFunction} directly without converting it to a 1D problem, thus allowing the inclusion of arbitrary regularization terms in the cost function. While in conventional 2D inversion SVD compression is mandatory, in multilinear inversion it is entirely optional, which leads to a flexible method that can handle non-separable kernels, include arbitrary regularization terms, and be easily extended to multidimensional NMR inversion.

The general mathematical framework for multilinear inversion was established by \cite{TensorInversion}. In our work, we use a simplified version of multilinear inversion and apply it to invert 2D and 3D NMR data. We impose nonnegativity constraints by using the projection gradient method \cite{bertsekas_goldstein_levitin_polyak_1976} instead of the BRD algorithm. The projection gradient method has the advantage that it can directly be applied to the tensor gradient. Accordingly, we first show how to calculate the tensor gradient for a cost function with zeroth-order Tikhonov regularization. Next, we show how to implement steepest descent for least-squares minimization while enforcing the nonnegativity constraint using the projection gradient method. Finally, we show how to include multi-order Tikhonov regularization terms, how to generalize to 3D NMR inversion, and how to deal with non-separable kernels and arbitrary regularization terms.
\subsection*{\normalfont{\bfseries{Tensor gradient with zeroth-order Tikhonov regularization}}}
We start by writing \eqref{eq:CostFunction} using Einstein's notation (i.e. double indices imply a sum) as
\begin{equation}
C(F)=\left((K_2)_{ik}F_{kq}(K_1)_{jq}-D_{ij}\right)^2.
\label{eq:CostEinsteinNotation}
\end{equation}
The $rs$-th component of the tensor gradient is then
\begin{align}
(\nabla C)_{rs} & = \frac{\partial}{\partial F_{rs}}  \left((K_2)_{ik}F_{kq}(K_1)_{jq}-D_{ij}\right)^2 & \nonumber \\
                & = 2 \left((K_2)_{ik}F_{kq}(K_1)_{jq}-D_{ij}\right)\cdot(K_2)_{i\alpha}(K_1)_{j\beta} \delta_{r\alpha}\delta_{s\beta} \nonumber \\
								&= 2 \left((K_2)_{ik}F_{kq}(K_1)_{jq}-D_{ij}\right)\cdot(K_2)_{ir}(K_1)_{js},
\label{eq:GradientEinsteinNotation}
\end{align}
where we have used $\partial F_{rs}/\partial F_{kq} = \delta_{rk}\delta_{sq} $. The tensor gradient can be written in matrix form as
\begin{equation}
(\nabla C) = 2 \tensor{K_2\tp} \tensor{E} \tensor{K_1},
\label{eq:TensorGradient}
\end{equation}
where we have defined the error $\tensor{E}$ as
\begin{equation}
\tensor{E}\equiv (\tensor{K_2}\tensor{F}\tensor{K_1\tp}-\tensor{D}).
\label{eq:TensorE}
\end{equation}
While the gradient in \eqref{eq:GradientK0} belongs to $\Re^{(N_1 \cdot N_2)\times 1}$, the gradient in \eqref{eq:TensorGradient} belongs to $\Re^{N_1 \times N_2}$. After a careful inspection, we can see that there are two advantages of using the tensor gradient form \eqref{eq:TensorGradient} instead of the augmented gradient one. 

\begin{table}[htb]
\footnotesize
\centering
 \begin{tabular}{c c c c c c} 
 \hline
 $N_{echo}$ & $N_{T_W}$& $N_{T_1}\times N_{T_2}$  & $m_A$ (MB)& $m_{TG}$ (MB) & $m_A/m_{TG}$  \\ [0.5ex] 
 \hline\hline
  300  &30 &  $50\times 50$ &    $180$  & $0.22$ &  804 \\[0.5ex] 
  300  &30 & $100\times 100$ &    $720$  & $0.42$ & 1731 \\ 
  300  &30 & $150\times 150$ &  $1,620$  & $0.65$ & 2500 \\ 
 5000  &30 &  $50\times 50$ &  $3,000$  & $3.23$ &  929 \\ 
 5000  &30 & $100\times 100$ & $12,000$  & $5.30$ & 2263 \\ 
 5000  &30 & $150\times 150$ & $27,000$  & $7.42$ & 3641 \\[1ex] 
 \hline
 \end{tabular}
\caption[1 Computation Memory]{Amount of memory required to compute the augmented gradient $\nabla(C(\tensor{F_A}))$ and the tensor gradient $\nabla(C(\tensor{F}))$. $N_{T_W}$  and  $N_{T_1}$ are the number of polarization steps $T_{W}$ and number of $T_1$ inversion points, respectively; $N_{echo}$ and $N_{T_2}$ are the number of echoes and number of $T_2$ inversion points, respectively; and $m_A$ and $m_{TG}$ denote the memory in MB required to calculate the augmented gradient and the tensor gradient (assuming 8 bytes per number).}
\label{tab:ComputingMemory}
\end{table}

\begin{figure}[!htb]
	\centering
	\subfloat[]{
		\includegraphics{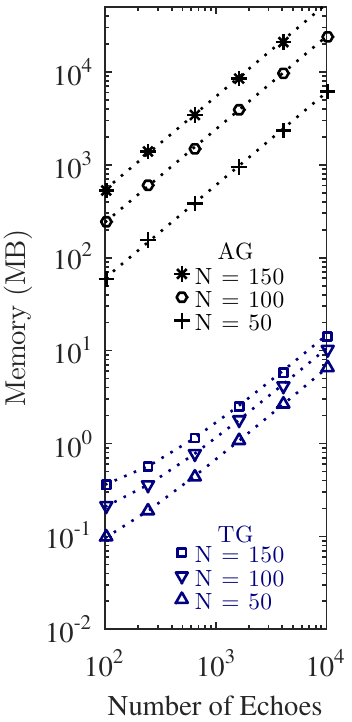}
		\label{fig:MatrixSize}}
	\quad
		\subfloat[]{
		\includegraphics{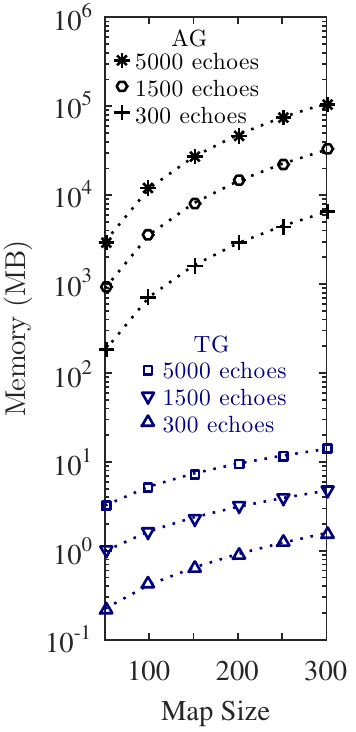}
		\label{fig:MatrixMapSize}}
	\caption[02 Memory comparison for 2D inversion]{Memory comparison for 2D inversion. Memory vs (a) number of echoes and (b) map size, required to calculate the augmented gradient (AG) and tensor gradient (TG) of cost function \eqref{eq:CostFunction}. In both (a) and (b), the maps are assumed to be symmetric ($N=N_{T_1}=N_{T_2}$). The number of polarization steps is 30 ($N_{T_W}=30$). We assume 8 bytes per number.}
	\label{fig:MatrixSizeTime}
\end{figure}

First and most importantly, the memory used by the tensor gradient can be several orders of magnitude less compared with the augmented gradient. For example, to compute the augmented gradient in \eqref{eq:GradientK0}, the augmented matrix $\tensor{K_A}$, $\tensor{F_A}$, and $\tensor{D_A}$ must be stored. This requires storing 
\begin{equation}
m_A=M_1 N_1 M_2 N_2 + M_1 M_2 + N_1 N_2
\label{eq:MemoryAugmentedMatrix2D}
\end{equation}
numbers in memory. In contrast, to compute the tensor gradient in \eqref{eq:TensorGradient}, we only need to store $\tensor{K_1}$, $\tensor{K_2}$, $\tensor{F}$, and $\tensor{D}$, whose total number of elements is
\begin{equation}
m_{TG}=M_1 N_1 + M_2 N_2 + M_1 M_2 + N_1 N_2.
\label{eq:MemoryMatrixGradient2D}
\end{equation}

Table \ref{tab:ComputingMemory} shows the memory requirements, $m_A$ and $m_{TG}$, to store the required matrices for the computation of the augmented gradient and the tensor gradient, respectively; $m_A$ and $m_{TG}$ have been calculated for 2 different number of echoes ($N_{echo}=300$ and $N_{echo}=$ 5000) and 3 different $T_1$-$T_2$ map sizes ($N_{T_1}\times N_{T_2}=50 \times 50$, $100 \times 100$, and $150 \times 150$). For a low resolution $T_1$-$T_2$ map of 50$\times$50 points obtained with 300 echoes and 30 polarization steps, the augmented gradient uses 804 times more memory than the tensor gradient. And, if we were to use 5000 echoes with 30 polarization steps to produce a $T_1$-$T_2$ map of $150\times150$ points, calculating the augmented gradient would require over 3600 times more memory than calculating the tensor gradient. Figures \ref{fig:MatrixSize} and \ref{fig:MatrixMapSize} show the amount of memory required to compute the augmented gradient and tensor gradient as a function of number of echoes and map size, respectively, with 30 $T_1$ polarization steps ($N_{T_W}=30$). For Figure \ref{fig:MatrixSize}, three different $T_1$-$T_2$ map sizes ($N_{T_1}\times N_{T_2}$) are considered: $50 \times 50$, $100 \times 100$, and $150 \times 150$, while for Figure \ref{fig:MatrixMapSize},  three different echo numbers ($N_{echo}$) are considered: 5000, 1500, and 300. For the same number of echoes and map size, the computation of the augmented gradient requires nearly 3 orders of magnitude more memory compared to the tensor gradient.

\begin{figure}[!htb]
	\centering
	\subfloat[]{
		\includegraphics{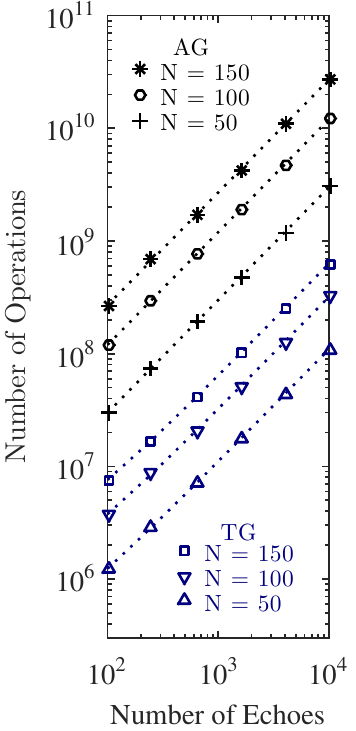}
		\label{fig:MatrixTime}}
	\quad
		\subfloat[]{
		\includegraphics{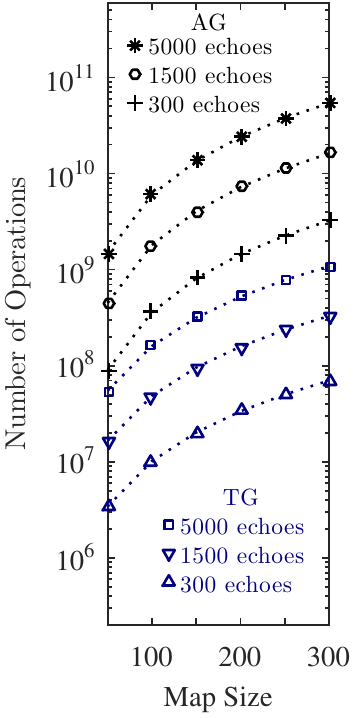}
		\label{fig:MatrixMapTime}}
	\caption[03 Number of operations comparison for 2D inversion]{Number of operations comparison for 2D inversion. Number of operations vs (a) number of echoes and (b) map size, required to calculate the augmented gradient (AG) and tensor gradient (TG) of cost function \eqref{eq:CostFunction}. In both (a) and (b), the maps are assumed to be symmetric ($N=N_{T_1}=N_{T_2}$). The number of polarization steps is 30 ($N_{T_W}=30$). We assume 8 bytes per number.}
	\label{fig:MatrixMapSizeTime}
\end{figure}

\begin{table}[htb]
\footnotesize
\centering
 \begin{tabular}{c c c c c c} 
 \hline
  $N_{echo}$ & $N_{T_W}$& $N_{T_1}\times N_{T_2}$ & $N_A$ & $N_{TG}$ & $N_{A}/N_{TG}$  \\ [0.5ex] 
 \hline\hline
  300  & 30 &  $50\times50$  & $9.00\EE{7}$  & $3.45\EE{6}$ & 26   \\ [0.5ex] 
  300  & 30 & $100\times100$  & $3.60\EE{8}$  & $1.02\EE{7}$ & 35     \\ 
  300  & 30 & $150\times150$  & $8.10\EE{8}$  & $2.03\EE{7}$ & 40     \\ 
 5000  & 30 &  $50\times50$  & $1.50\EE{9}$  & $5.52\EE{7}$ & 27     \\ 
 5000  & 30 & $100\times100$  & $6.00\EE{9}$  & $1.60\EE{8}$ & 37     \\ 
 5000  & 30 & $150\times150$  & $1.35\EE{10}$ & $3.16\EE{8}$ & 43   \\  [1ex] 
 \hline
 \end{tabular}
\caption[2 Computation Operations]{Number of operations (products and additions) needed to compute the augmented gradient $\nabla(C(\tensor{F_A}))$ and the tensor gradient $\nabla(C(\tensor{F}))$. $N_{echo}$ and $N_{T_W}$ are the number of echoes and polarization steps, respectively; $N_{T_1}$ and $N_{T_2}$  are the number of $T_1$ and $T_2$ inversion points, respectively; and $N_A$ and $N_{TG}$ are the number of operations needed to compute the augmented gradient and the tensor gradient, respectively.}
\label{tab:ComputingOperations}
\end{table}

Second, the total number of operations (products and additions), $N_A$, that it takes to compute the augmented gradient is 
\begin{equation}
N_A = 4 M_1 N_1 M_2 N_2,
\label{eq:NumberOperationsAugmentedGradient}
\end{equation}
while the total number of operations needed to compute the tensor gradient is 
\begin{equation}
N_{TG} = 2(M_1 N_1 M_2 + M_1 N_1 N_2 + M_1 M_2 N_2 + N_1 M_2 N_2),
\label{eq:NumberOperationsTensorGradient}
\end{equation}
where we have neglected terms of order $O(N^2)$. Table \ref{tab:ComputingOperations} shows the total number of operations used to calculate the augmented gradient ($N_A$) and tensor gradient ($N_{TG}$). We observe that the ratio $N_A/N_{TG}$ does not change as much as $m_A/m_{TG}$ as the number of echoes increase. Thus, the main advantage of using tensor gradients is the efficient use of memory, which translates into faster memory access times and, therefore, a faster algorithm. Figures \ref{fig:MatrixTime} and \ref{fig:MatrixMapTime} show the number of operations required to compute the augmented gradient and tensor gradient as a function of number of echoes and map size, respectively, with 30 $T_1$ polarization steps ($N_{T_W}=30$). For Figure \ref{fig:MatrixMapTime}, three different echo numbers ($N_{echo}$) are considered: 5000, 1500, and 300, while for Figure \ref{fig:MatrixTime}, three different $T_1$-$T_2$ map sizes ($N_{T_1}\times N_{T_2}$) are considered: $50 \times 50$, $100 \times 100$, and $150 \times 150$. The number of operations required to compute the augmented matrix is larger by a factor roughly equal to the smallest of $N_{echo}$, $N_{T_W}$, $N_{T_1}$, and $N_{T_2}$.

To include zeroth-order Tikhonov regularization, we add the penalty term $\alpha^2 \left\| \tensor{F} \right\|_F^2$ to the cost function,
\begin{equation}
C(\tensor{F})=\left\| \tensor{K_2}\tensor{F}\tensor{K_1\tp} - \tensor{D} \right\|_F^2 + \alpha^2 \left\| \tensor{F} \right\|_F^2.
\label{eq:CostFunctionZerothOrderReg}
\end{equation}
The tensor gradient then takes the form
\begin{equation}
(\nabla C) = 2 \tensor{K_2\tp} \tensor{E} \tensor{K_1} + 2 \alpha^2 \tensor{F}.
\label{eq:TensorGradientZerothOrderReg}
\end{equation}
In the case of 3D NMR inversion, the difference in memory requirements is even greater. For example, \cite{arns2007multidimensional} considered the 3D NMR sequence $T_2$-$D$-$DG_0^2$, where without compression, they found that the uncompressed augmented matrix would require 67 TB (\cite{arns2007multidimensional} cites 61 TB because they assumed that 1 MB = $2^{20}$ bytes, while we assume that 1 MB = $10^6$ bytes). Even after kernel compression using SVD, they still require 2 GB of memory. By contrast, even without kernel compression, the multilinear inversion method uses only 216 MB to store the kernels. In other words, using the full kernels, the multilinear inversion method requires $3.1\EE{5}$ times less memory than the uncompressed kernels, and 9.9 times less memory than the fully compressed kernels. 

\begin{figure}[!htb]
	\centering
	\subfloat[]{
		\includegraphics{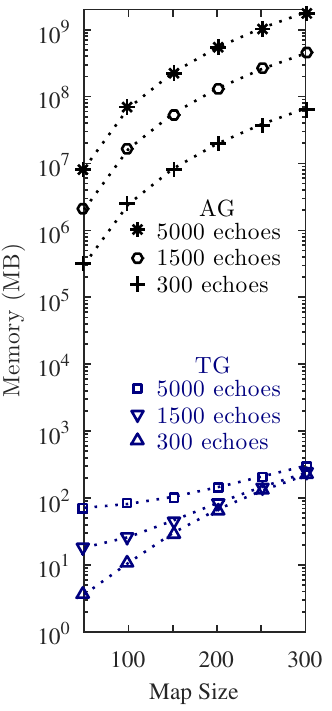}
		\label{fig:Memory3DEcho}}
	\quad
	\subfloat[]{
		\includegraphics{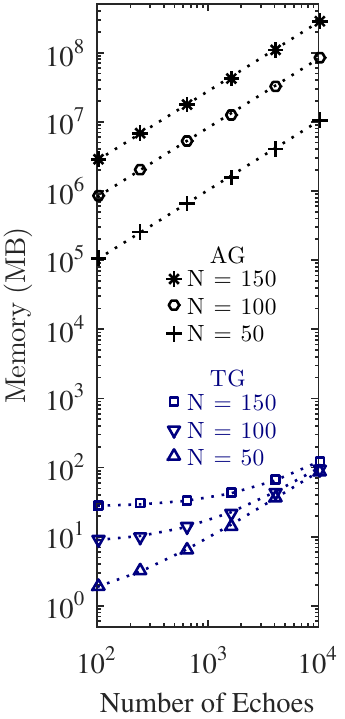}
		\label{fig:Memory3DMap}}
	\caption[04 Memory comparison for 3D inversion]{Memory comparison for 3D inversion. Memory vs (a) map size and (b) number of echoes, required to calculate the augmented gradient (AG) and tensor gradient (TG) of a 3D NMR cost function \cite{arns2007multidimensional} with separable kernels. In both (a) and (b), the maps are assumed to be symmetric ($N=N_1=N_2=N_3$), $M_2=32$, and $M_3=32$. We assume 8 bytes per number.}	
	\label{fig:figureMemory3D}
\end{figure}

In this work, we will consider the 3D NMR sequence $D$-$T_1$-$T_2$. The number of elements required to calculate the augmented gradient of a 3D NMR cost function with separable kernels is 
\begin{equation}
m_A=M_1 N_1 M_2 N_2M_3N_3 + M_1M_2M_3+N_1N_2N_3,
\label{eq:MemoryAugmentedMatrix3D}
\end{equation}
while for the tensor gradient we have
\begin{equation}
m_{TG}=M_1 N_1 + M_2 N_2 + M_3N_3 + M_1M_2M_3+N_1N_2N_3.
\label{eq:MemoryMatrixGradient3D}
\end{equation}
Figure \ref{fig:figureMemory3D} shows plots of $m_A$ and $m_{TG}$ as a function of maps size and number of echoes. If we were to use the same number of parameters as \cite{arns2007multidimensional} ($N_D=N_{T_1}=N_{T_2}=100$, $N_{echo}=8196$, $N_g=32$, and $N_{T_W}=32$), we would require 67 TB of memory to compute the augmented gradient. On the other hand, the tensor gradient would only require 82 MB, i.e., $8.21\times 10^5$ times less memory. 

\subsection*{\normalfont{\bfseries{Multilinear minimization}}}
We use the steepest descent method to minimize the cost function $C(\tensor{F})$ by moving the current $\ell$-th solution $\tensor{f}_\ell$ along a descent direction $\tensor{d}_\ell/||\tensor{d}_\ell||_F$, namely,
\begin{equation}
\tensor{f}_{\ell+1} = \tensor{f}_{\ell} + \gamma_\ell \tensor{d}_\ell/||\tensor{d}_\ell||_F,
\label{eq:NewStep}
\end{equation}
where $\gamma_\ell > 0$ is the step-size and $\tensor{d}_\ell/||\tensor{d}_\ell||_F$ is the search direction given by the negative of the gradient,
\begin{equation}
\tensor{d}_{\ell} = - (\tensor{K_2\tp} \tensor{E} \tensor{K_1} + \alpha^2 \tensor{F}),
\label{eq:SteepestDescent}
\end{equation}
where we have omitted the inconsequential factor of 2. The step-size $\gamma_\ell$ is obtained by minimizing the line-search scalar function $\mathcal{L}(\gamma_\ell)$ defined by
\begin{equation}
 \mathcal{L}(\gamma_\ell) = C(\tensor{P}(\tensor{f}_{\ell} + \gamma_\ell \tensor{d}_\ell/||\tensor{d}_\ell||_F)),
\label{eq:LineSearchFunction}
\end{equation}
where $\tensor{P}$ is the projection gradient operator. Its action on the $kq$-th component of $\tensor{f}$ is defined as

\begin{equation}
\tensor{P}(f_{kq}) = \begin{cases}
f_{kq} & \text{if } f_{kq}> 0\\
0  &\text{otherwise}.
\end{cases}
\end{equation}

If the new step $\tensor{f}_{\ell+1}$ falls in a region where some of its components are negative, the projection operator sets these components to zero. There are several ways to minimize the line search scalar function $\mathcal{L}(\gamma_\ell)$ \cite{chouzenoux2010efficient}. For its simplicity, we evaluate $\mathcal{L}(\gamma_\ell)$ at a predetermined set ($\Gamma$) of step-size values ($\gamma_i$), and choose the step-size ($\gamma_\ell$) for which $\mathcal{L}$ is minimum. The step-size list is made of 20 logarithmically (base 10) spaced values from $1\EE{-7}$ to $1\EE{0}$. In practice, this interval will depend on the nature of the data. The minimization is stopped when the fractional change of the cost function $(C(\tensor{f_{\ell}})-C(\tensor{f_{\ell+1}}))/C(\tensor{f_{\ell+1}})$ is less than a specified tolerance (tol), usually set to $1\EE{-3}$. Algorithm \ref{alg:BilinearSolver} describes the nonnegative multilinear inversion method for two dimensions.

\begin{algorithm}
\caption{Nonnegative Bilinear Steepest Descent}\label{euclid}
\label{alg:BilinearSolver}
\begin{algorithmic}[1]
\Require{$\Gamma$=$\{\gamma_i\}$; $\tensor{f}_0$, (initial guess); $\tensor{K_1}$; $\tensor{K_2}$; $\tensor{D}$; tol (tolerance)}
\Procedure{BilinearSolver}{$\Gamma$, $\tensor{f}_0$, $\tensor{K_1}$, $\tensor{K_2}$, $\tensor{D}$, tol}
\While{ $(C(\tensor{f_{\ell}})/C(\tensor{f_{\ell+1}})-1)>$ tol } 
\State $\tensor{e_\ell} \gets (\tensor{K_2}\tensor{f}_\ell\tensor{K_1\tp}-\tensor{D})$ \Comment{current error}
\State $\tensor{d}_\ell \gets -(\tensor{K_2\tp} \tensor{e}_\ell \tensor{K_1} + \alpha^2 \tensor{f}_\ell) $ \Comment{current step direction}
\State $\gamma_{\ell} \gets \argmin\limits_{\gamma_i}  C(\tensor{P}(\tensor{f}_{\ell} + \gamma_i \tensor{d}_\ell/||\tensor{d}_\ell||_F))$ \Comment{best step-size}
\State $\tensor{f}_{\ell+1} \gets \tensor{P}(\tensor{f}_{\ell}+\gamma_\ell \tensor{d}_\ell/||\tensor{d}_\ell||_F) $ \Comment{update solution}
\EndWhile
\EndProcedure
\end{algorithmic}
\end{algorithm}

\subsection*{\normalfont{\bfseries{Generalization}}}
We show how to generalize the cost function to include multi-order Tikhonov regularization in two and three dimensions with separable kernels, how to handle non-separable kernels, and how to include arbitrary regularization terms.
\subsubsection*{\normalfont{\bfseries{ 1) Two-dimensional cost function with zeroth- and second- order Tikhonov regularization}}}
Second-order Tikhonov regularization imposes constraints on the smoothness of the solution. To add multi-order Tikhonov regularization, we write the cost function as
\begin{equation}
C(\tensor{F})=\left\| \tensor{K_2}\tensor{F}\tensor{K_1\tp} - \tensor{D} \right\|_F^2 + \alpha_0^2 \left\| \tensor{F} \right\|_F^2 +\alpha_2^2 \left( \left\| \tensor{D_L^{(2)}F} \right\|_F^2 + \left\| \tensor{FD_R^{(2)}} \right\|_F^2 \right),
\label{eq:CostFunctionMultiOrderTikhonov}
\end{equation}
where $\tensor{D_L^{(2)}}$ is the left-acting second-order derivative operator that acts on the rows of $\tensor{F}$, $\tensor{D_R^{(2)}}$ is the right-acting second-order derivative operator that acts on the columns of $\tensor{F}$, and $\alpha_0$ and $\alpha_2$ are the zeroth- and second-order regularization coefficients. The tensor gradient is then given by
\begin{equation}
(\nabla C) = 2 \tensor{K_2\tp} \tensor{E} \tensor{K_1} + 2 \alpha_0^2 \tensor{F} +2\alpha_2^2 \left( \tensor{D_L^{(2)T}} \tensor{D_L^{(2)}} \tensor{F} + \tensor{F}\tensor{D_R^{(2)T}} \tensor{D_R^{(2)}}   \right),
\label{eq:GradientMultiOrderTikhonov}
\end{equation}
where $\tensor{D_L^{(2)T}}$ and $\tensor{D_R^{(2)T}}$ denote the transpose matrices, $\tensor{E}$ is defined in \eqref{eq:TensorE}. The representation of  $\tensor{D_L^{(2)}}$  and  $\tensor{D_R^{(2)}} $ that we use is (in the central scheme):

\begin{eqnarray}
(\tensor{D_L^{(2)}})_{mnqr}F_{qr}&=&\delta_{nr}(\delta_{m+1,q}-2\delta_{mq}+\delta_{m-1,q})F_{qr},  \\
(\tensor{D_R^{(2)}})_{mnqr}F_{qr}&=&\delta_{mq}(\delta_{n+1,r}-2\delta_{nr}+\delta_{n-1,r})F_{qr}.  \\
\label{eq:2DDerivativeRepresntation}\nonumber
\end{eqnarray}
\subsubsection*{\normalfont{\bfseries{ 2) Three-dimensional cost function with zeroth- and second-order Tikhonov regularization}}}
For the case of 3D NMR inversion with separable kernels, the cost function with zeroth- and second-order regularization has the form
\begin{align}
C(\tensor{F})  & = \|(K_1)_{iq} (K_2)_{jr} (K_3)_{ks}F_{qrs}-D_{ijk}\|^2_F+\alpha_0^2 \|F_{ijk}\|^2_F& \nonumber \\
               & + \alpha_2^2 \|\tensor{D}^{(2)} _1 \tensor{F}\|^2_F+ \alpha_2^2 \|\tensor{D}^{(2)}_2 \tensor{F}\|^2_F + \alpha_2^2 \|\tensor{D}^{(3)}_3 \tensor{F}\|^2_F,
\label{eq:CostFunctionThreeDimensionalMultiOrder}
\end{align}
where $\tensor{D}^{(2)}_1$, $\tensor{D}^{(2)}_2$, and $\tensor{D}^{(2)}_3$ are the second-order discrete partial-differential operators that act on the first, second, and third components of $\tensor{F}$, respectively. In the central scheme, they are given by
\begin{eqnarray}
(\tensor{D}^{(2)}_1 )_{mnpqrs}F_{qrs}&=&\delta_{nr}\delta_{ps}(\delta_{m+1,q}-2\delta_{m,q}+\delta_{m-1,q})F_{qrs},  \\
(\tensor{D}^{(2)}_2 )_{mnpqrs}F_{qrs}&=&\delta_{ps}\delta_{mq}(\delta_{n+1,r}-2\delta_{n,r}+\delta_{n-1,r})F_{qrs},  \\
(\tensor{D}^{(2)}_3 )_{mnpqrs}F_{qrs}&=&\delta_{mq}\delta_{nr}(\delta_{p+1,s}-2\delta_{p,s}+\delta_{p-1,s})F_{qrs}. \\
\label{eq:DiscreteSecondOrderPartialDifferential}\nonumber
\end{eqnarray}
The multilinear-gradient is then given by
\begin{eqnarray}
	(\nabla C(\tensor{F}))_{mnp} &=& 2 \tensor{(K_1\tp K_1)}_{mq} \tensor{(K_2\tp K_2)}_{nr} \tensor{(K_3\tp K_3)}_{ps} F_{qrs} \nonumber \\
	 &-&2D_{ijk} \tensor{(K_1)}_{im} \tensor{(K_2)}_{jn} \tensor{(K_3)}_{kp} +2\alpha_0^2 F_{mnp} \nonumber \\
	&+&2\alpha_2^2 (\tensor{D}^{(2)}_1)_{ijkmnp}(\tensor{D}^{(2)}_1)_{ijkqrs}F_{qrs} \nonumber \\
	&+&2\alpha_2^2 (\tensor{D}^{(2)}_2)_{ijkmnp}(\tensor{D}^{(2)}_2)_{ijkqrs}F_{qrs} \nonumber \\
	&+&2\alpha_2^2 (\tensor{D}^{(2)}_3)_{ijkmnp}(\tensor{D}^{(2)}_3)_{ijkqrs}F_{qrs}. \nonumber \\
\label{eq:GradientThreeDimensional}
\end{eqnarray}
\subsubsection*{\normalfont{\bfseries{ 3) Two-dimensional cost function with non-separable kernels and arbitrary regularization term}}}
The 2D cost function for non-separable kernels with an arbitrary regularization term has the form
\begin{equation}
C(\tensor{F})=\| K_{ijkl}F_{kl} -D_{ij}\|_F^2 +\zeta(\tensor{F}) ,
\label{eq:CostFunctionNonSeparableKernel}
\end{equation}
where $K_{ijkl}$ represents a non-separable kernel. The tensor gradient is then given by
\begin{equation}
(\nabla C)_{rs} = 2 K_{ijrs} \left(K_{ijkl}F_{kl} -D_{ij} \right)+ \frac{\partial \zeta(\tensor{F})}{\partial F_{rs}}.
\label{eq:GradientNonSeparableKernel}
\end{equation}

%% file: Results.tex
\section{Results}
\label{sec:Results}
Inverting with a given regularization parameter $\alpha$ and finding the optimal $\alpha$ itself are two different problems. During the inversion process the value of the regularization term is kept constant, therefore we emphasize just on the inversion process and not on how to choose the optimal regularization parameter for which plenty of methods already exist \cite{mitchell_numerical_2012,day2011inversion,hansen1998rank}. 

All inversions were performed with MATLAB\textsuperscript{\textregistered} R2015a on a core i7-5820K CPU @ 3.30 GHz with 32 GB of RAM. And all measurements were carried out with an Oxford Instruments GeoSpec2 NMR spectrometer working at 2.17 MHz. Table \ref{tab:Summary} shows a summary of the inversion parameters and results for examples 2) to 5).
\subsection*{\normalfont{\bfseries{ 1) Two-dimensional NMR inversion with zeroth-order regularization}}}
We invert a laboratory $T_1$-$T_2$ NMR data set using zeroth-order Tikhonov regularization. The cost function we seek to minimize is given by \eqref{eq:CostFunctionZerothOrderReg}. The $T_1$-$T_2$ sequence is an IR-CPMG sequence of Berea sandstone saturated with deionized water. The data consists of 46,296 echoes with 30 polarization steps $T_W$ with a signal-to-noise ratio of 60 and an inter-echo time of 0.108 ms.

The performance of the bilinear inversion method is compared with two other linear solvers with nonnegative constraints: the Lawson-Hanson (LH) method \cite{lawson1974solving}, and the Butler-Reeds-Dawson (BRD) method. We used the \textit{lsqnonneg} function of MATLAB\textsuperscript{\textregistered} R2015a as our implementation of the Lawson-Hanson method. We implemented the Butler-Reeds-Dawson method following \cite{LalithaSolvingFredholm2002}. While the bilinear inversion algorithm is memory efficient and can handle all of the data, the other two methods quickly consume memory as the size of the matrices increases. Therefore, \textit{only} for comparison purposes, we bin each $T_2$ decay into 300 points. Binning the data reduces the number of echoes, not the number of polarization times. 

The cost function we seek to minimize is 
\begin{equation}
C_{Bil}(\tensor{F})=\left\| \tensor{K_2}\tensor{F}\tensor{K_1\tp} - \tensor{D} \right\|_F^2 + \alpha^2 \left\| \tensor{F} \right\|_F^2.
\label{eq:CostFunctionZerothOrderReg2}
\end{equation}
The bilinear steepest descent method can minimize the cost function \eqref{eq:CostFunctionZerothOrderReg2} directly. However, to apply the LH or BRD method, we must transform \eqref{eq:CostFunctionZerothOrderReg2} into the equivalent 1D inversion problem, namely, 
\begin{equation}
C_{BRD}(\tensor{F_A})=\left\| \tensor{K_A}\tensor{F_A}- \tensor{D_A} \right\|_F^2 + \alpha^2\left\| \tensor{F_A}\right\|_F^2 ,
\label{eq:CostFunctionOneD2}
\end{equation}
where $\tensor{K_A}=\tensor{K_1}\otimes\tensor{K_2}$, $\tensor{F_A}=\text{vec}(\tensor{F})$, and $\tensor{D_A}=\text{vec}(\tensor{D})$. The BRD method can now be used to minimize \eqref{eq:CostFunctionOneD2}. In order to use the LH method, \eqref{eq:CostFunctionOneD2} must be written in the standard form \cite{elden1977algorithms}
\begin{equation}
C_{LH}(\tensor{F_A})= \left\| \left(\begin{array}{c} \tensor{K_A}\\ \alpha \textbf{I}  \end{array}\right)\tensor{F_A} - \left(\begin{array}{c} \tensor{D_A}\\ \mathbf{0} \end{array}\right) \right\|_F^2,
\label{eq:StandardForm}
\end{equation}
where $\textbf{I}$ is an identity matrix with the same number of columns as $\tensor{K_A}$, and $\mathbf{0}$ is a zero row-vector with the same number of rows as $\textbf{I}$. The LH method can now be used to minimize the cost function \eqref{eq:StandardForm}.

\begin{figure*}[!htb]
\center
\subfloat[]{\label{fig:alpha1Map50}\includegraphics{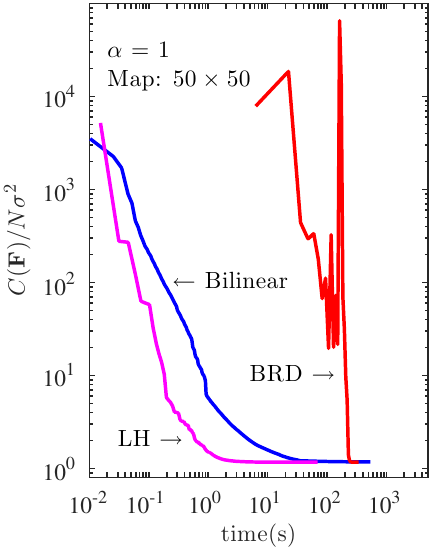}}
\subfloat[]{\label{fig:alpha1Map100}\includegraphics{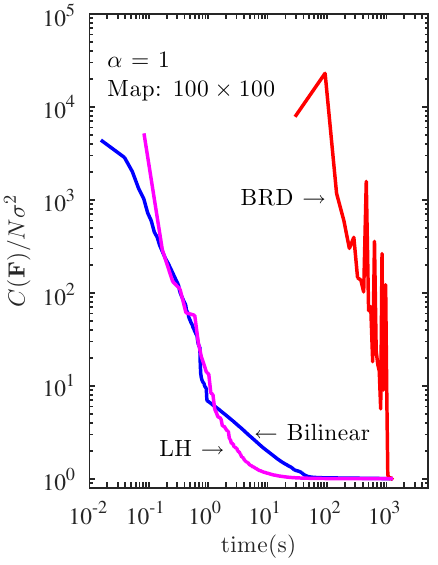}}
\subfloat[]{\label{fig:alpha1Map150}\includegraphics{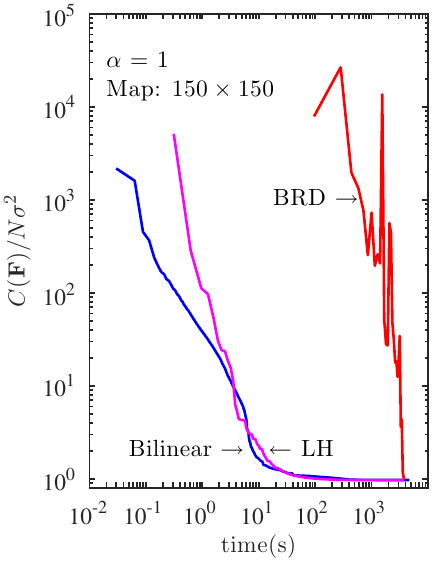}}
\vfill
\subfloat[]{\label{fig:alpha10Map50}\includegraphics{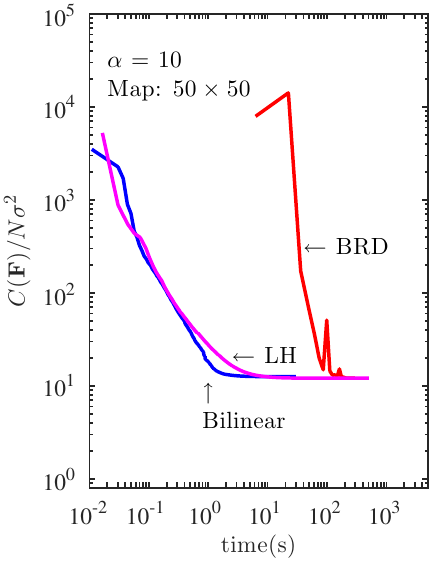}}
\subfloat[]{\label{fig:alpha10Map100}\includegraphics{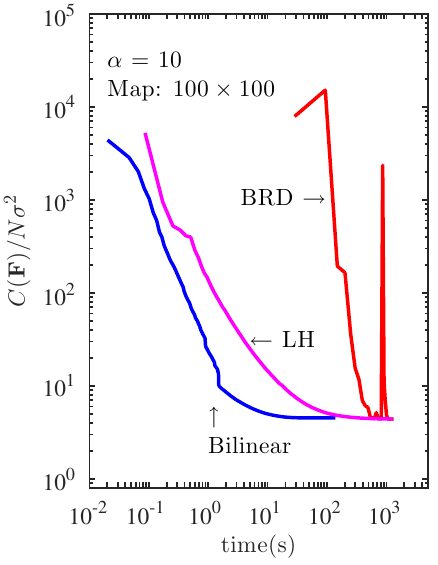}}
\subfloat[]{\label{fig:alpha10Map150}\includegraphics{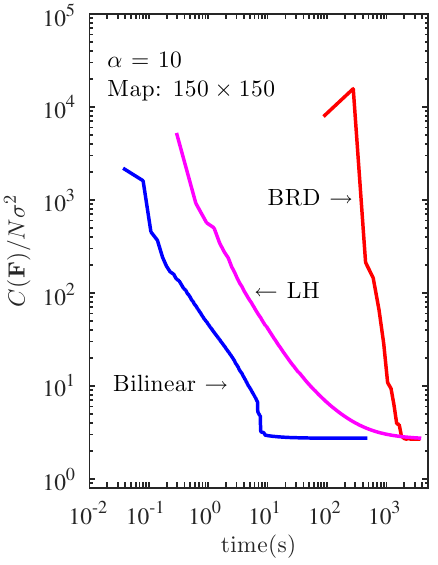}}
\caption[05 2D NMR inversion time comparison]{2D NMR inversion time comparison. Cost function minimization time of a $T_1$-$T_2$ inversion of water-saturated Berea sandstone for the Butler-Reeds-Dawson, Lawson-Hawson, and Bilinear inversion methods. The value of the cost function is plotted as a function of minimization time for different map sizes and zeroth-order regularization parameters. The $T_1$-$T_2$ maps are symmetric ($N_{T_1}=N_{T_2}$). For all cases, the data is binned to 300 data points ($N=300$), the number of $T_W$ steps is 30, and $\sigma$ is the standard deviation of the noise.}\label{fig:MinimizationTimes}
\end{figure*}

\begin{figure*}[!htb]
\center
\subfloat[]{\label{fig:MAP_Bilinear_Map_N300_M_100_alpha1}\includegraphics{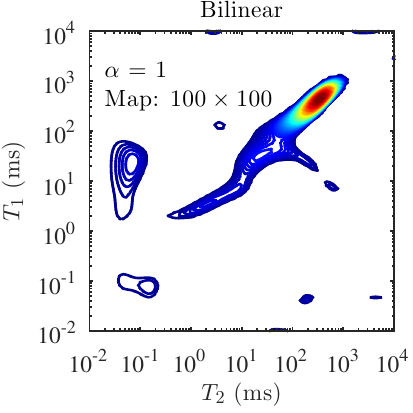}}
\subfloat[]{\label{fig:MAP_Bilinear_Map_N300_M_150_alpha1}\includegraphics{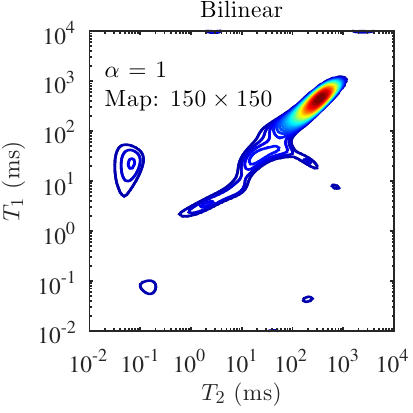}}
\subfloat[]{\label{fig:MAP_Bilinear_Map_N300_M_100_alpha10}\includegraphics{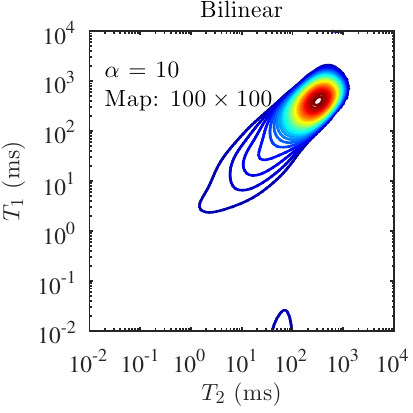}}
\subfloat[]{\label{fig:MAP_Bilinear_Map_N300_M_150_alpha10}\includegraphics{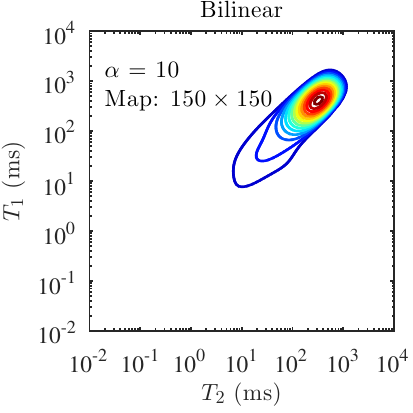}}
\vfill
\subfloat[]{\label{fig:MAP_lsqnonneg_Map_N300_M_100_alpha1}\includegraphics{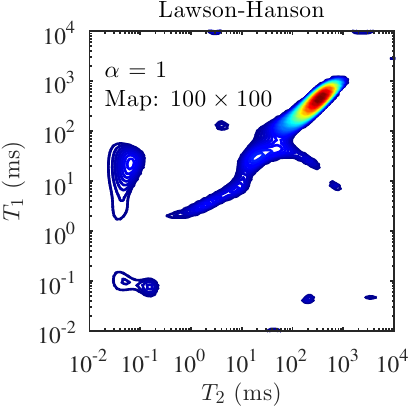}}
\subfloat[]{\label{fig:MAP_lsqnonneg_Map_N300_M_150_alpha1}\includegraphics{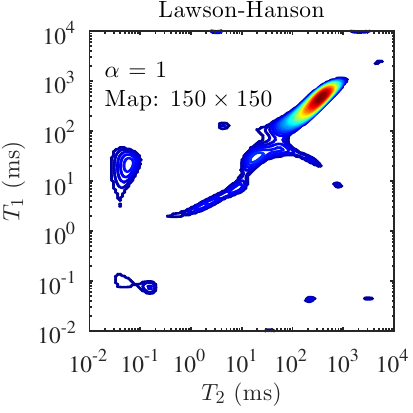}}
\subfloat[]{\label{fig:MAP_lsqnonneg_Map_N300_M_100_alpha10}\includegraphics{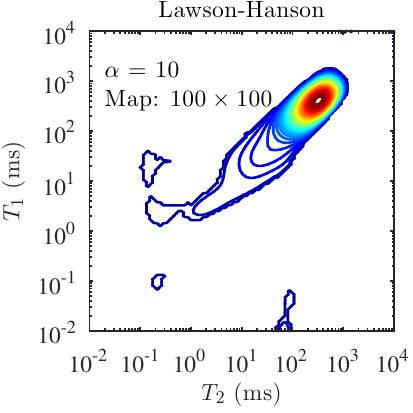}}
\subfloat[]{\label{fig:MAP_lsqnonneg_Map_N300_M_150_alpha10}\includegraphics{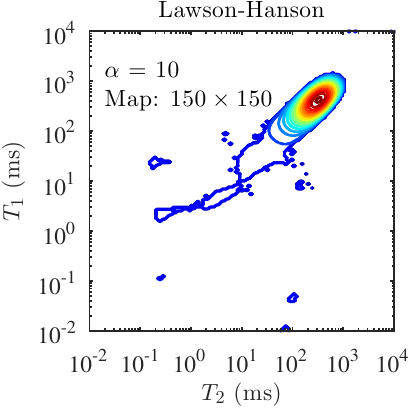}}
\vfill
\subfloat[]{\label{fig:MAP_BRD_Map_N300_M_100_alpha1}\includegraphics{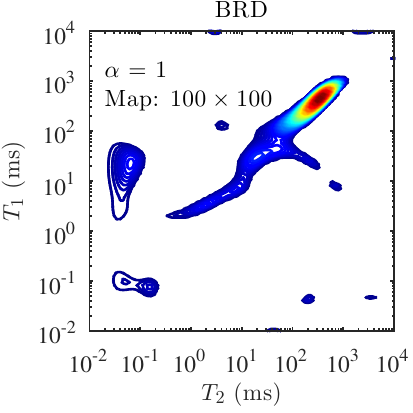}}
\subfloat[]{\label{fig:MAP_BRD_Map_N300_M_150_alpha1}\includegraphics{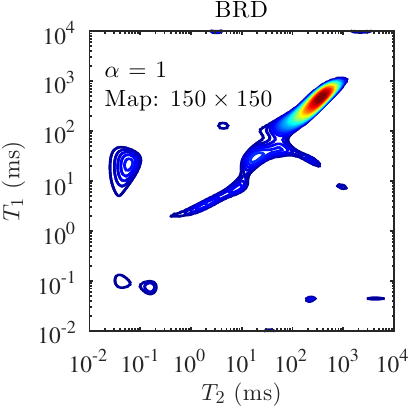}}
\subfloat[]{\label{fig:MAP_BRD_Map_N300_M_100_alpha10}\includegraphics{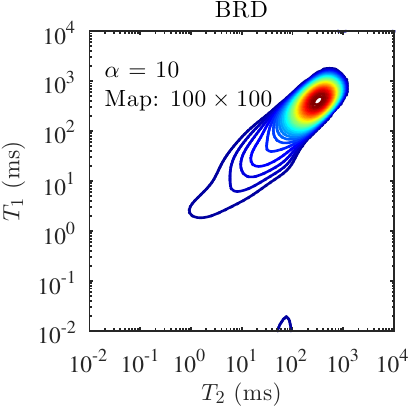}}
\subfloat[]{\label{fig:MAP_BRD_Map_N300_M_150_alpha10}\includegraphics{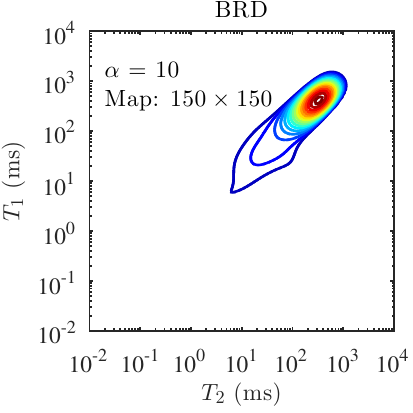}}
\caption[06 2D NMR inversion comparison]{2D NMR inversion comparison. Comparison of the $T_1$-$T_2$ 2D distributions obtained with the LH, BRD, and bilinear inversion methods for different map sizes and zeroth-order regularization parameters. The $T_1$-$T_2$ maps are symmetric ($N_{T_1}=N_{T_2}$). (a-d) $T_1$-$T_2$ distributions obtained with Bilinear inversion. (e-h) $T_1$-$T_2$ distributions obtained with the Lawson-Hanson method. (i-l) $T_1$-$T_2$ distributions obtained with the Butler-Reeds-Dawson method. In all cases, the number of binned echoes is $N_{echo,bin}=300$, and the number of polarization steps is 30 ($N_{T_W}=30)$.}\label{fig:InversionMaps}
\end{figure*}

Figure \ref{fig:MinimizationTimes} shows the elapsed time in minimizing the normalized cost functions $C_{Bil}/N\sigma^2$, $C_{LH}/N\sigma^2$, and $C_{BRD}/N\sigma^2$ corresponding to the bilinear, LH, and BRD methods for three $T_1$-$T_2$ map sizes ($N_{T_1}=N_{T_2}=50,100,150$) and two different values of regularization parameter ($\alpha = 1,10$). Each data set is binned to 300 echoes ($N=N_{echo,bin}=300$) with 30 polarization steps ($N_{T_W}=30$), comprising a total of 9,000 data points. Minimizations are stopped after 1 hour of computation (or earlier if the minimization converges). 

Figure \ref{fig:MinimizationTimes} shows that the BRD curve is not monotonically decreasing. This is because the BRD method (Venkataramanan's implementation) is not directly minimizing the cost function $C_{BRD}(\tensor{F})/N\sigma^2$, but an effective cost function $\chi(\tensor{c})$ that automatically imposes the nonnegativity constraint \cite{LalithaSolvingFredholm2002}. 

In all cases of Figure \ref{fig:MinimizationTimes}, the BRD method is 200 to 300 times slower than the LH or the bilinear methods. For small map sizes ($50\times50$) and low values of regularization parameter ($\alpha=1$), the LH method is approximately 10 times faster than the bilinear method (Figure \ref{fig:alpha1Map50}). However, for a given value of $\alpha$, as the size of the map increases (Figures \ref{fig:alpha1Map50}-\ref{fig:alpha1Map150} and \ref{fig:alpha10Map50}-\ref{fig:alpha10Map150}) the speed of convergence of the bilinear method approaches that of the LH method, eventually surpassing it. For large values of $\alpha$ (Figures \ref{fig:alpha10Map50}, \ref{fig:alpha10Map100}, and \ref{fig:alpha10Map150}), the bilinear method converges 10 to 100 times faster than the LH method. 

Figure \ref{fig:InversionMaps} shows the resulting $T_1$-$T_2$ maps from the minimizations of Figures \ref{fig:alpha1Map100}, \ref{fig:alpha1Map150}, \ref{fig:alpha10Map100}, and \ref{fig:alpha10Map150}. One can see that the steepest descent bilinear inversion method produces results that are comparable to the ones produced by the BRD method, both of which are not only smoother than the maps produced by the LH method, but are more physically consistent with the condition $T_1>T_2$. 

In most cases of Figure \ref{fig:MinimizationTimes}, the BRD method performed considerably slower than the bilinear and the LH method. Therefore, in what remains of this work we will not consider the BRD method anymore. 

\begin{figure}[!htb]
	\centering
	\includegraphics{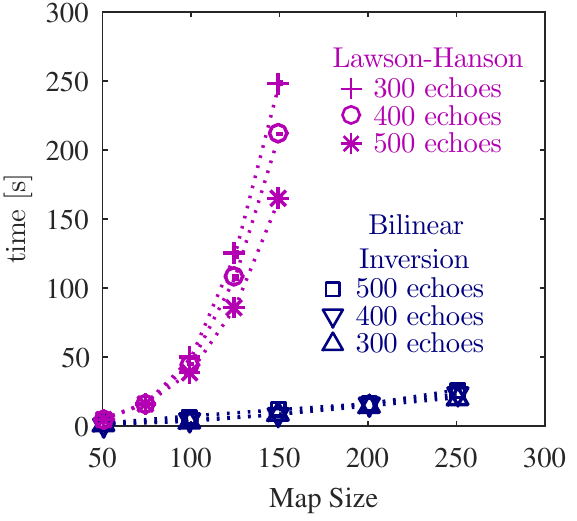}
	\caption[07 Lawson-Hanson vs bilinear inversion]{Lawson-Hanson vs bilinear inversion. Time elapsed to reach 1.01 times the minimum value of the cost function \eqref{eq:CostFunctionZerothOrderReg} with $\alpha=10$ as a function of map size for different number of echoes. The minimum value of the cost function is taken as the value it attains using a tolerance of at most $1\EE{-6}$ (see Algorithm \ref{alg:BilinearSolver}).}
	\label{fig:OperationsLHBilinear}
\end{figure}

Finally, Figure \ref{fig:OperationsLHBilinear} shows the time it takes the cost function \eqref{eq:CostFunctionZerothOrderReg} (with $\alpha=10$) to reach 1.01 times its minimum value with the bilinear and the LH methods as a function of the map size.  The minimum value of the cost function is taken as the value it attains using a tolerance of at most $1\EE{-6}$ (see Algorithm \ref{alg:BilinearSolver}). For the Lawson-Hanson method, only map sizes up to $150\times150$ were considered, as the memory requirements were beyond 32 GB.

\subsection*{\normalfont{\bfseries{ 2) Two-dimensional NMR inversion without kernel or data compression with zeroth-order regularization}}}
We obtained $T_1$-$T_2$ NMR data from water-saturated Berea sandstone using an IR-CPMG sequence. The data set consists of $N_{echo}=46,296$ echoes and $N_{T_W}=30$ polarization steps (a total of $1.38\EE{6}$ data points) with a signal-to-noise ratio of 60. We invert the $T_1$-$T_2$ data set by minimizing the cost function \eqref{eq:CostFunctionZerothOrderReg} with $\alpha=1$ and obtain a $T_1$-$T_2$ distribution with a resolution of $N_{T_1}=N_{T_2}=150$ (a $150\times150$ 2D map).
The memory required to calculate the cost function gradient if we were to use the LH method would be 250 GB. By contrast, the bilinear method requires 67 MB (3,738 times less memory). 

\begin{figure}[!htb]
	\centering
		\includegraphics{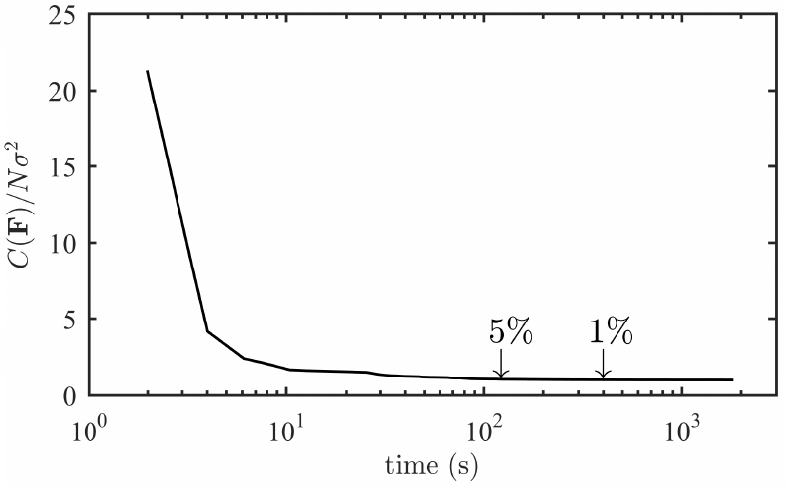}
	\caption[08 Full 2D inversion minimization time]{Full 2D inversion minimization time. The curve corresponds to the cost function \eqref{eq:CostFunctionZerothOrderReg} with $\alpha=1$ and a map size of $150\times150$ normalized by the number of echoes ($N$) and the variance of the noise ($\sigma^2$). The inverted data corresponds to a laboratory data set of water-saturated Berea sandstone with 46,296 echoes and 30 polarization steps (a total of $1.38\EE{6}$ data points and 22,500 fitting parameters). Bilinear inversion can minimize the cost function to 5 \% of its minimum value in under 120 seconds and to 1 \% of its minimum value in 400 seconds. The minimum value of the cost function is taken as the value it attains using a tolerance of at most $1\EE{-6}$ (see Algorithm \ref{alg:BilinearSolver}).}\label{fig:Figure_FullDataInversionCost}
\end{figure}

\begin{figure}[!htb]
	\centering
		\includegraphics{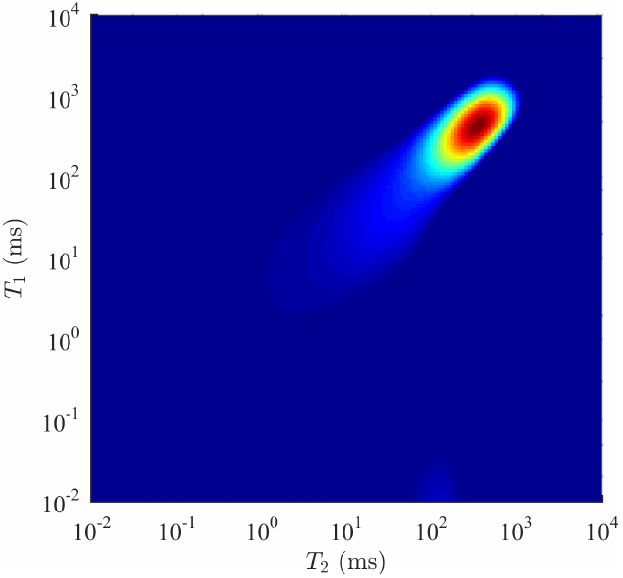}
	\caption[09 Full 2D inversion]{Full 2D inversion. $T_1$-$T_2$ distribution of water-saturated Berea sandstone obtained after inverting a data set corresponding to an IR-CPMG sequence with 46,296 echoes and 30 polarization steps (a total of $1.38\EE{6}$ data points and 22,500 fitting parameters) with zeroth-order regularization parameter ($\alpha=1$) using the bilinear steepest descent algorithm. The map resolution is $N_{T_1}\times N_{T_2}=150\times150$. The inversion is consistent with $T_1>T_2$. The blur around $T_2=10^2$ ms and $T_1=10^{-2}$ ms is due to a small overall data offset; its effect contributes to only 1.26 \% of the total signal.}
	\label{fig:Figure_FullDataInversionMap}
\end{figure}

We can minimize the cost function to 5\% of its minimum in under 120 seconds, and to 1\% of its minimum in 400 seconds. The minimum value is taken as the value the cost function attains after using a tolerance of at most $1\EE{-6}$ (see Algorithm \ref{alg:BilinearSolver}). Figure \ref{fig:Figure_FullDataInversionCost} shows the minimization curve of the cost function as a function of time. The resulting map, shown in Figure \ref{fig:Figure_FullDataInversionMap},  is consistent with $T_1>T_2$. The blur around $T_2=10^2$ ms and $T_1=10^{-2}$ ms is due to a small overall data offset; its effect contributes to 1.26 \% of the total signal.

\subsection*{\normalfont{\bfseries{ 3) Three-dimensional $D$-$T_1$-$T_2$ NMR inversion of synthetic data}}}
\begin{figure*}[!htb]
\center
\subfloat[]{\label{fig:SyntheticInitial_DT1T2map}\includegraphics[scale=0.25]{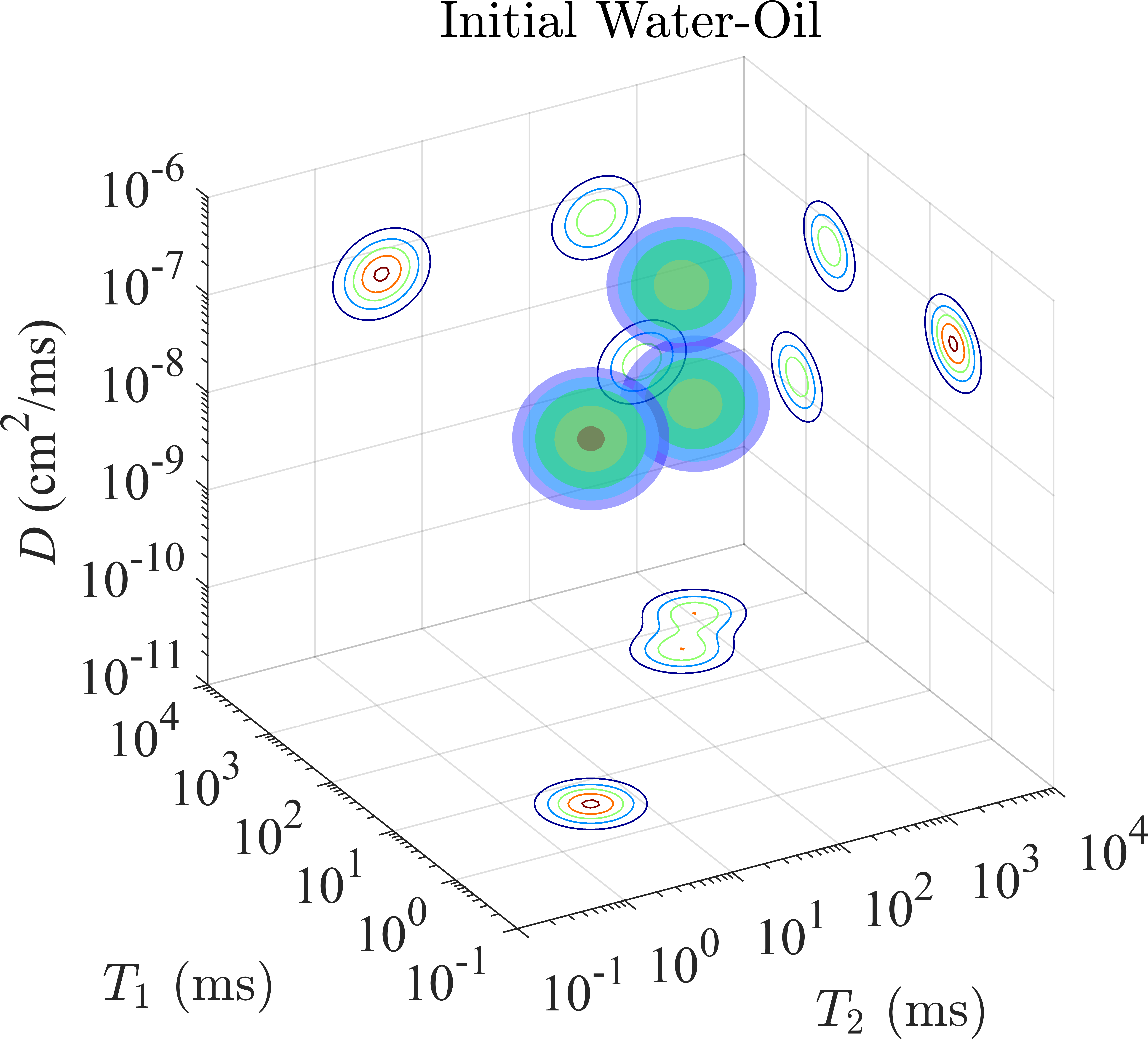}}
\subfloat[]{\label{fig:SyntheticInitial_T1T2map}\includegraphics{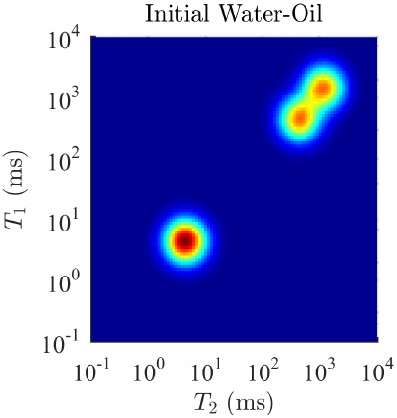}}
\subfloat[]{\label{fig:SyntheticInitial_DT2map}\includegraphics{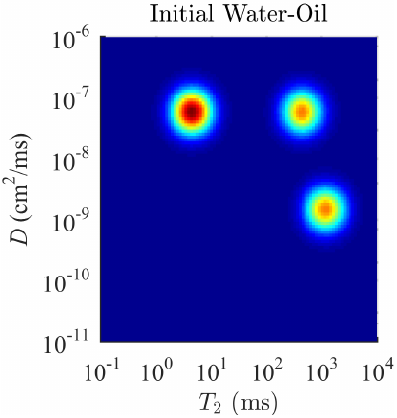}}
\subfloat[]{\label{fig:SyntheticInitial_DT1map}\includegraphics{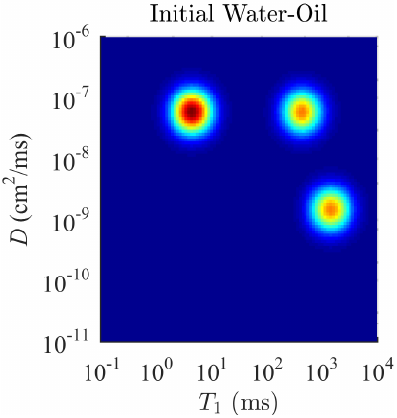}}
\vfill
\subfloat[]{\label{fig:SyntheticFinal_DT1T2map}\includegraphics[scale=0.25]{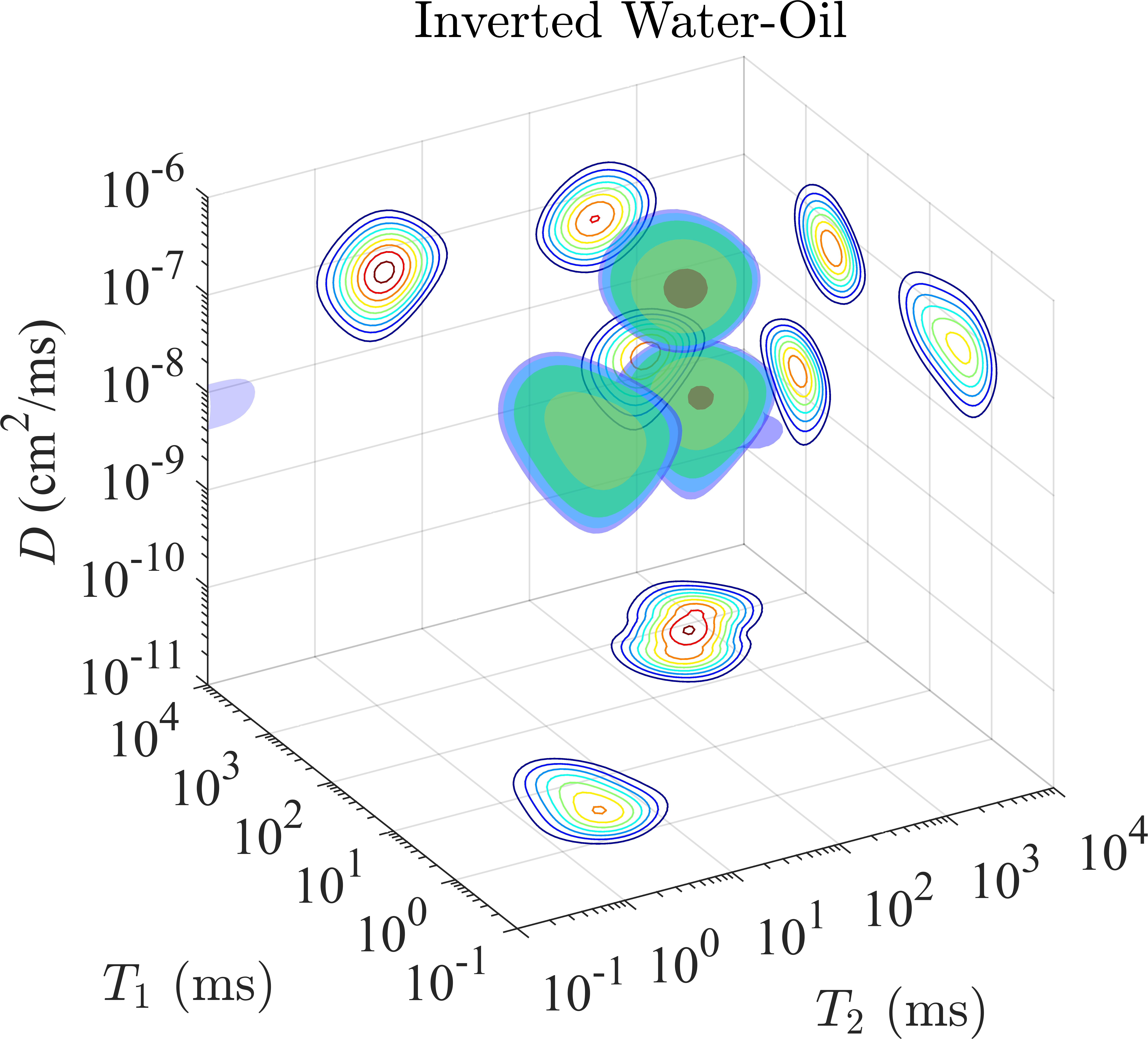}}
\subfloat[]{\label{fig:SyntheticFinal_T1T2map}\includegraphics{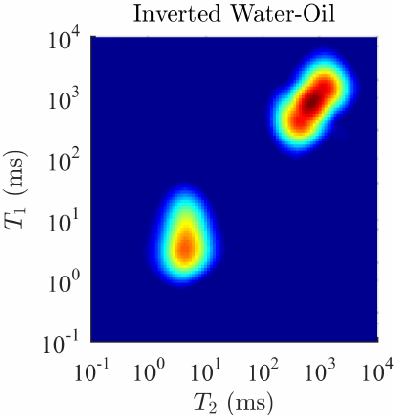}}
\subfloat[]{\label{fig:SyntheticFinal_DT2map}\includegraphics{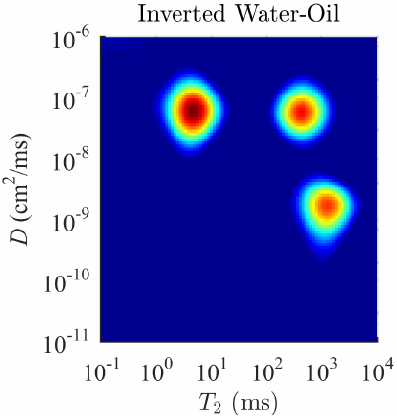}}
\subfloat[]{\label{fig:SyntheticFinal_DT1map}\includegraphics{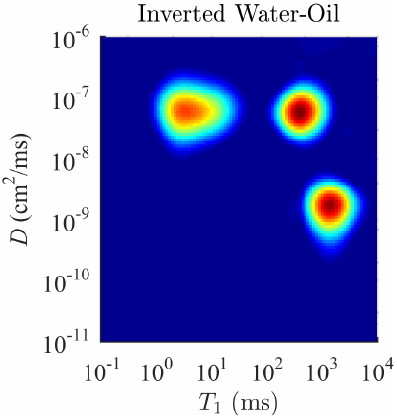}}
\caption[10 3D NMR inversion of a synthetic model]{3D NMR inversion of a synthetic model. (a) Original water-oil 3D $D$-$T_1$-$T_2$ synthetic model, and (b-d) its 2D projections. The synthetic model corresponds to a mixture of bound water, free water, and light oil in a 4:3:3 saturation ratio. Table \ref{tab:SyntheticModel} shows the fluid properties used to construct the synthetic model. The resolution of the 3D $D$-$T_1$-$T_2$ distribution is $N_{T_1}\times N_{T_2}\times N_D=100\times100\times100$. The synthetic $D$-$T_1$-$T_2$ distribution is used to simulate NMR decay data with 1\% Gaussian noise. Multilinear inversion is applied to the synthetic time series to yield (e) a new $N_{T_1}\times N_{T_2}\times N_D=100\times100\times100$ inverted 3D $D$-$T_1$-$T_2$ distribution, and (f-h) its corresponding 2D projections. The normalized $\chi^2$ of the difference between the data predicted by the initial and inverted synthetic model is $\chi^2=1.3$, while the correlation between the initial and inverted 3D $D$-$T_1$-$T_2$ distribution is 0.91. }\label{fig:Synthetic_Map}
\end{figure*}

We use synthetic data to quantify the accuracy of the nonnegative steepest descent multilinear inversion method for the case of 3D NMR inversion. First, we start with a 3D $D$-$T_1$-$T_2$ distribution corresponding to a synthetic mixture of free water, bound water, and light oil. We assume log-normal $D$, $T_1$, and $T_2$ NMR distributions. Table \ref{tab:SyntheticModel} describes the NMR properties of the synthetic mixture. The total $D$-$T_1$-$T_2$ distribution is constructed by adding the contribution of the $D$-$T_1$-$T_2$ distribution of each phase. The individual $D$-$T_1$-$T_2$ distributions are formed by taking the tensor product of the $D$, $T_1$, and $T_2$ distributions of each phase. All the distributions are assumed to be log-normal. Figures \ref{fig:SyntheticInitial_DT1T2map}, \ref{fig:SyntheticInitial_T1T2map}, \ref{fig:SyntheticInitial_DT2map}, and \ref{fig:SyntheticInitial_DT1map} show the initial synthetic 3D $D$-$T_1$-$T_2$ NMR distribution, the $T_1-T_2$, the $D-T_2$, and the $D-T_1$ projection maps, respectively. The initial $D$-$T_1$-$T_2$ distribution has a resolution of $N_D=100$, $N_{T_1}=100$, and $N_{T_2}=100$. 

\begin{figure}[!htb]
	\centering
		\includegraphics{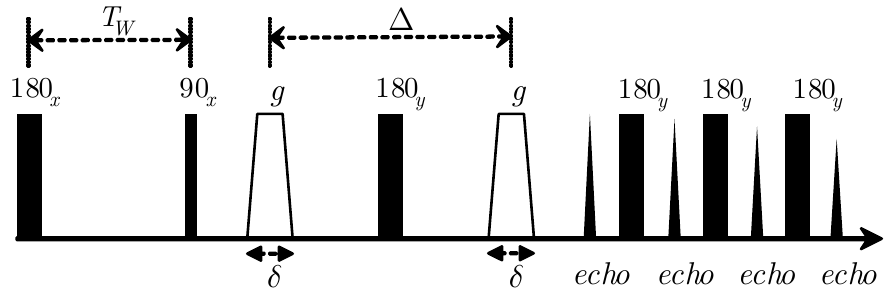}
	\caption[11 3D pulse sequence]{3D pulse sequence. The sequence starts with an inversion recovery period during which the spins are allowed to relax longitudinally for a time period $T_W$, followed by a PFGSE sequence, and ends with a CPMG echo train acquisition sequence. $g$ denotes the amplitude magnetic field gradient pulse, $\delta$ is the duration of the magnetic field gradient pulse, and $\Delta$ is the time between the magnetic field gradient pulses.}
	\label{fig:3DPulseSequence}
\end{figure}

Next, we calculate data time series using 
\begin{equation}
S(t,T_W,g)=S_0\iiint{e^{-\gamma^2 g^2 \delta^2 (\Delta-\delta/3)D} e^{-t/\tau_2} (1-2e^{-T_W/\tau_1}) f(\tau_2,\tau_1,D) d\tau_1 d\tau_2 dD},
\label{eq:DT1T2signal}
\end{equation}
which corresponds to an IR-PFGSE-CPMG sequence, shown in Figure \ref{fig:3DPulseSequence} \cite{Stejskal1964}. We use 14 polarization steps ($N_{T_W}=14$) and 31 gradient steps ($N_g=31$) with 10,000 echoes per sequence with an inter-echo spacing of 1 ms. After adding 1\% Gaussian noise, we bin the data into 302 points ($N_{echo,bin}=302$), reducing the number of data points to $N_{data}=131,068$. 

We then invert the synthetic data by minimizing the cost function \eqref{eq:CostFunctionThreeDimensionalMultiOrder} with $\alpha_0 = 0.05$ and $\alpha_2 = 0$. Figure \ref{fig:SyntheticFinal_DT1T2map} shows the $D$-$T_1$-$T_2$ distribution obtained after 2,263 seconds of computation. Figures \ref{fig:SyntheticFinal_T1T2map}, \ref{fig:SyntheticFinal_DT2map}, and \ref{fig:SyntheticFinal_DT1map} show the corresponding $T_1-T_2$, the $D-T_2$, and the $D-T_1$ projection maps, respectively. 

Multilinear inversion used 9.3 MB of memory, or $1.124\times 10^5$ times less memory than the 1049 GB needed by the LH method. Table \ref{tab:SyntheticModelInverted} shows the estimated fluid properties after inversion. The normalized $\chi^2$ of the difference between the data predicted by the initial and inverted synthetic model is $\chi^2=1.3$, while the correlation between the initial and final 3D distribution is 0.91. 

\begin{table}[htb]
\footnotesize
\centering
 \begin{tabular}{c c c c} 
 \hline
  phase & $T_1$, $T_2$ (ms) & $D$ (cm$^2$/ms)  & saturation  \\ [0.5ex] 
 \hline
  bound water  & 4.30  &    $6.00\times10^{-8}$     & 0.400 \\[0.5ex] 
  free water   & 464   &     $6.00\times10^{-8}$      & 0.300 \\ 
  light oil    & 1500  &     $1.50\times10^{-9}$      & 0.300 \\[1ex] 
 \hline
 \end{tabular}
\caption[3 Synthetic model original]{Fluid properties used to construct a synthetic 3D $D$-$T_1$-$T_2$ distribution of bound water, free water, and light oil. $T_1$ and $T_2$ are the longitudinal and transverse relaxation rates, respectively, and $D$ is the diffusion coefficient.}
\label{tab:SyntheticModel}
\end{table}


\begin{table}[htb]
\footnotesize
\centering
 \begin{tabular}{c c c c c c c c} 
 \hline
	phase & $T_{1}$ (ms) & error & $T_{2}$ (ms) & error & $D$ (cm$^2$/ms)   & error & saturation  \\ [0.5ex] 
 \hline
  bound water  & 3.8 & -11\%   & 4.32  & +0.5\% &    $5.85\times10^{-8}$ & -2.5\%   & 0.408 \\[0.5ex] 
  free water   & 391 & -16\%  & 422   & -9\%&    $5.50\times10^{-8}$   & -8\% & 0.295 \\ 
  light oil    & 1394 & -7\% & 1262   & -16\%&    $1.65\times10^{-9}$  & +10\% & 0.297 \\[1ex] 
 \hline
 \end{tabular}
\caption[5 Synthetic model result]{Estimated fluid properties obtained from the inversion of the three-phase synthetic model. The resulting values for $T_1$, $T_2$, and $D$ are the peak values extracted from the 2D projections of the 3D $D$-$T_1$-$T_2$ distribution. The saturations are calculated from the 2D projections by adding the partial porosities corresponding to each peak. The errors shown are relative errors with respect to the initial value.}
\label{tab:SyntheticModelInverted}
\end{table}

\subsection*{\normalfont{\bfseries{ 4) Three-dimensional $D$-$T_1$-$T_2$ NMR inversion of a mixture of water and heavy water.}}}
The laboratory sample, referred to as sample 38, consists of a solution of 2\% NaCl and $<$ 1\% CuSO$_4$ in a 18.30 ml mixture of H$_2$O and D$_2$O. To obtain $D$-$T_1$-$T_2$ NMR data from sample 38 we used an IR-PFGSE-CPMG sequence (see Figure \ref{fig:3DPulseSequence}). The kernel for such sequence is given by,
\begin{equation}
K(t,T_{W},g;\tau_2,\tau_1,D)= e^{-t/\tau_2}e^{-\gamma^2 g^2 \delta^2 (\Delta-\delta/3)D} (1-2e^{-T_W/\tau_1}),
\label{eq:KernelDT1T2}
\end{equation}
where $\tau_1$ and $\tau_2$ are the $T_1$ and $T_2$ relaxation times, $t$ denotes time, $T_W$ is the polarization time, $\gamma$ is the proton gyromagnetic ratio, $g$ is the magnetic field gradient, $\delta$ is the magnetic field gradient duration, and $\Delta$ is time between magnetic field gradient pulses. In this test we used $\Delta=30$ ms, $\delta=5$ ms, and an inter-echo time $T_E$ of 1 ms. Table \ref{tab:Sample38SequenceNMRproperties} shows the NMR properties of sample 38.

\begin{figure*}[!htb]
\center
\subfloat[]{\label{fig:Sample38_DT1T2map}\includegraphics[scale=0.25]{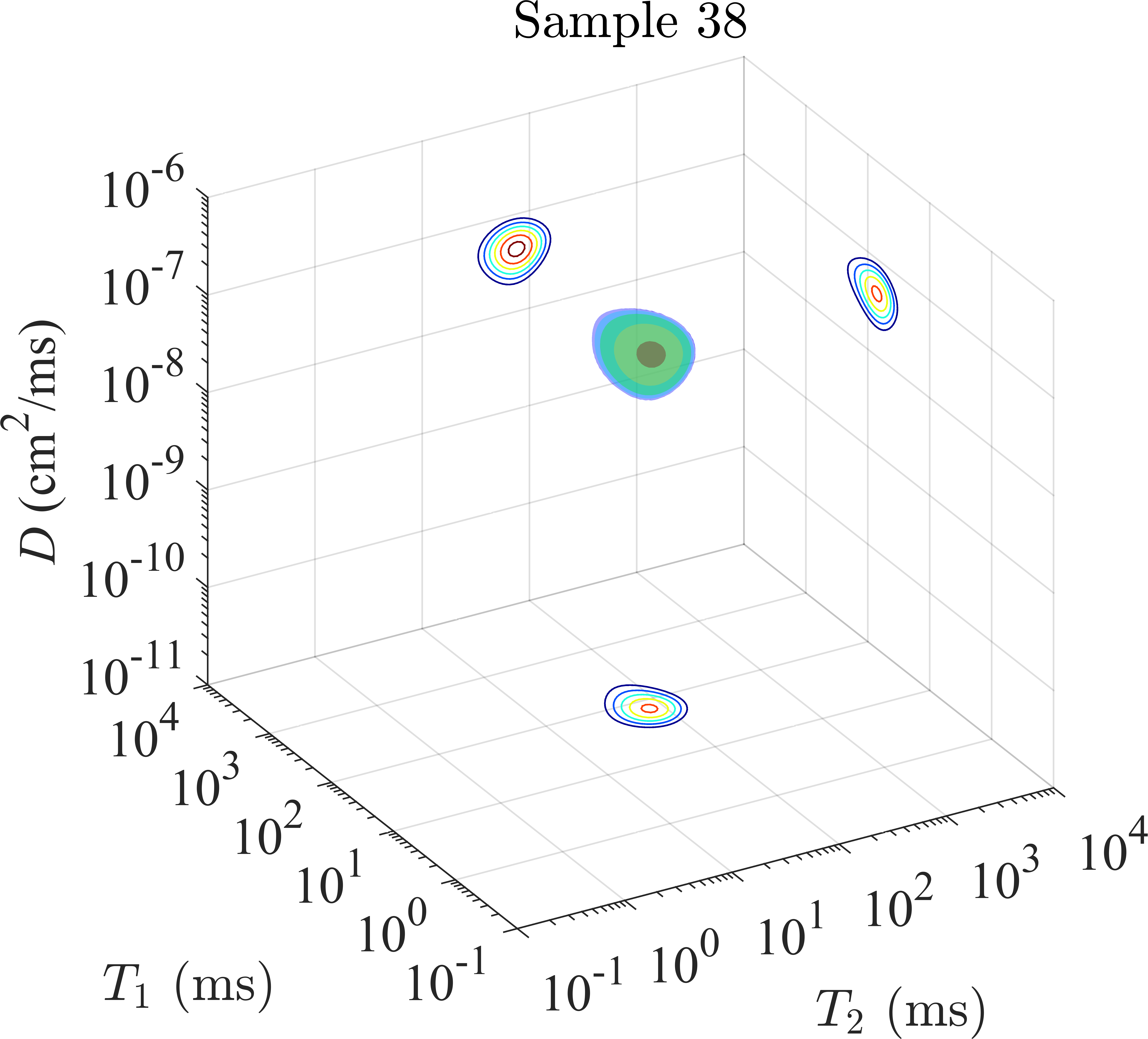}}
\subfloat[]{\label{fig:Sample38_T1T2map}\includegraphics{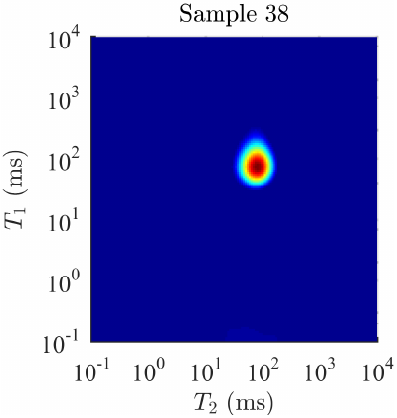}}
\subfloat[]{\label{fig:Sample38_DT2map}\includegraphics{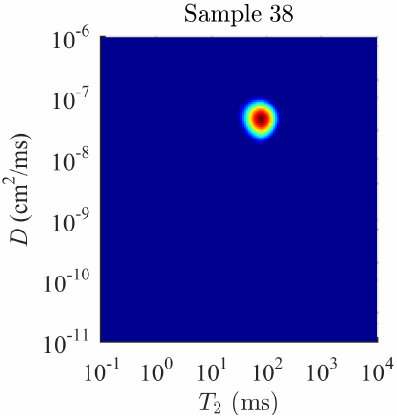}}
\subfloat[]{\label{fig:Sample38_DT1map}\includegraphics{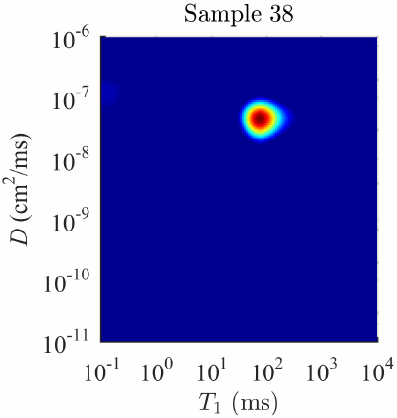}}
\caption[12 3D NMR inversion of sample 38]{3D NMR inversion of sample 38. (a) 3D $D$-$T_1$-$T_2$ distribution of sample 38, and (b-d)  its 2D projections. The resolution of the 3D $D$-$T_1$-$T_2$ distribution is $N_{T_1} \times N_{T_2} \times  N_D=150\times150\times150$. It was obtained after 5,955 seconds of computation time from a data set of 103 echoes $N_{echo,bin}=103$, with 12 polarization steps ($N_{T_W}=12$), and 11 magnetic field gradient steps ($N_g=11$) (for a total of 13,596 data points and $3.375\times10^6$ fitting parameters). The $\chi^2$ of the fit to the full data ($N_{echo}=512$) is 1.30. The signal to noise ratio was 87.}\label{fig:Sample38_Inversion_map}
\end{figure*}

The data set consists of $N_{echo}=512$ echoes with $N_{T_W}=12$ polarization steps and $N_g=11$ gradient steps with a signal-to-noise ratio of 87. Before inverting, we bin the number of echoes into $N_{echo,bin}=103$ points without affecting the polarization or gradient steps. We then invert the $D$-$T_1$-$T_2$ data set by minimizing the cost function \eqref{eq:CostFunctionThreeDimensionalMultiOrder} with $N_{T_{2}}=N_{T_{1}}=N_{D}=150$, $\alpha_0=2$, and $\alpha_2=0.1$. Figure \ref{fig:Sample38_Inversion_map} shows the inverted 3D $D$-$T_1$-$T_2$ distribution of sample 38 obtained after 5,955 seconds of inversion. Multilinear inversion used 27.3 MB of memory, or $1.34\times10^4$ less memory than the 367 GB needed by the LH method. The normalized $\chi^2$ of the difference between the experimental data and the one predicted by the inversion is $\chi^2=1.30$.

\begin{table}[htb]
\footnotesize
\centering
 \begin{tabular}{c c c c c} 
 \hline\hline
       & $T_{1}$ (ms) & $T_{2}$ (ms) & $D$ (cm$^2$/ms)    & H.I.  \\ [0.5ex] 
 \hline
       & 80           &   80         & $4.2\times10^{-8}$ &0.424  \\[0.5ex] 
 \hline
 \end{tabular}
\caption[5 Sample-38 experimental parameters ]{ Sample-38 $D$-$T_1$-$T_2$ NMR properties. The sample consists of a solution of 2\% NaCl and $<$ 1\% CuSO$_4$ in a 18.30 ml mixture of H$_2$O and D$_2$O. $T_1$, $T_2$, $D$, and H.I. are the longitudinal and transverse relaxation rates, diffusion constant, and hydrogen index of sample 38, respectively.}
\label{tab:Sample38SequenceNMRproperties}
\end{table}

\subsection*{\normalfont{\bfseries{ 5) Three-dimensional $D$-$T_1$-$T_2$ NMR inversion of a laboratory data set of water-saturated Berea sandstone}}}

As our final example, we show the 3D $D$-$T_1$-$T_2$ NMR inversion of a laboratory data set for water-saturated Berea sandstone. The sequence used to obtain the $D$-$T_1$-$T_2$ distribution is an IR-PFGSE-CPMG pulse sequence (see Figure \ref{fig:3DPulseSequence}), with $\Delta=5$ ms, $\delta=3$ ms, and an inter-echo time of $T_E=0.4$ ms. The NMR data obtained consists of $N_{echo}=10,240$ echoes, with 13 polarization steps ($N_{T_W}=13$), and 8 magnetic field gradient steps ($N_g=8$) with a signal-to-noise ratio of 187. Before inversion, the data was binned to reduce the number of echoes to 300.

\begin{figure*}[!htb]
\center
\subfloat[]{\label{fig:BereaSS_DT1T2map}\includegraphics[scale=0.25]{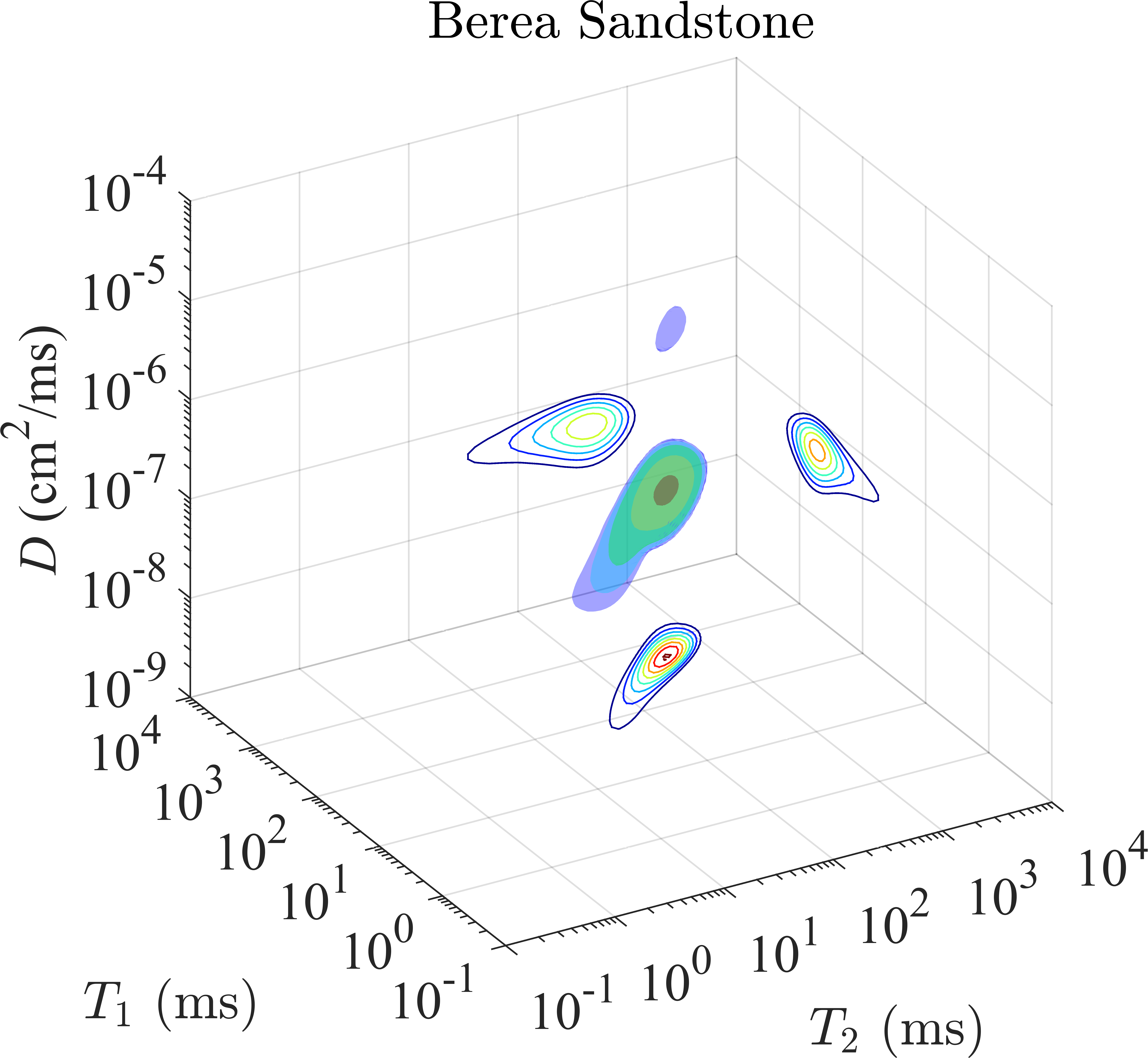}}
\subfloat[]{\label{fig:BereaSS_T1T2map}\includegraphics{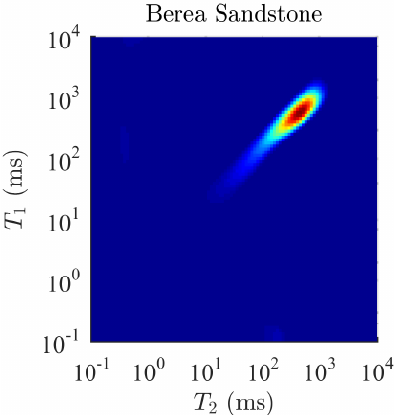}}
\subfloat[]{\label{fig:BereaSS_DT2map}\includegraphics{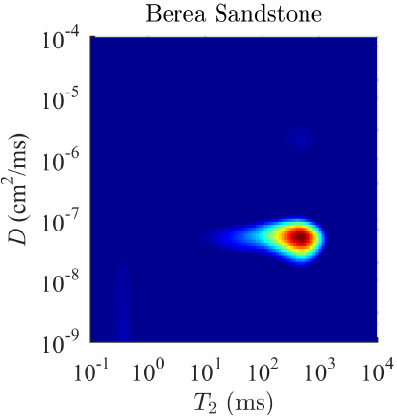}}
\subfloat[]{\label{fig:BereaSS_DT1map}\includegraphics{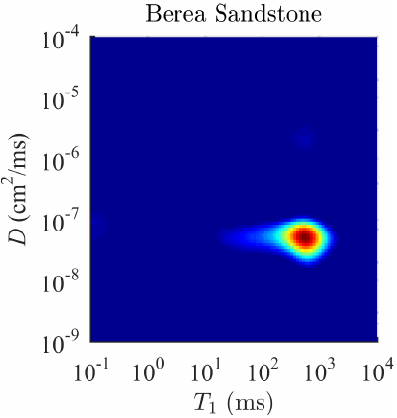}}
\caption[13 3D NMR inversion of Berea sandstone]{3D NMR inversion of Berea sandstone. (a) $D$-$T_1$-$T_2$ 3D distribution of water-saturated Berea sandstone, and (b-d) its 2D projections. The resolution of the 3D $D$-$T_1$-$T_2$ distribution is $N_{T_1}\times N_{T_2}\times N_D=100\times100\times100$. It was obtained after 5,184 seconds of computation time from a data set of 300 echoes $N_{echo,bin}=300$, with 13 polarization steps ($N_{T_W}=13$), and 8 magnetic field gradient steps ($N_g=8$) (for a total of 31,200 data points and $10^6$ fitting parameters). The $\chi^2$ of the fit to the full data ($N_{echo}=10,240$) is 1.00. The signal to noise ratio was 187.}
\label{fig:BereaSS_Inversion_map}
\end{figure*}

We invert the data by minimizing the cost function \eqref{eq:CostFunctionThreeDimensionalMultiOrder} with $\alpha_0=0.1$ and $\alpha_2=0$. Figure \ref{fig:BereaSS_Inversion_map} shows the inverted 3D $D$-$T_1$-$T_2$ distribution of water-saturated Berea sandstone after 5,184 seconds of computation time. Multilinear inversion used 8.5 MB of memory, or $2.93\times10^4$ less memory than the 250 GB needed by the LH method. The normalized $\chi^2$ of the difference between the experimental data and the one predicted by the inversion is $\chi^2=1.00$.

\begin{table}[!htb]
\scriptsize
\centering
 \begin{tabular}{c c c c c} 
 \hline\hline
             & Berea Sandstone & Synthetic Model & Sample 38       & Berea Sandstone \\
  Inversion Type   & $T_1$-$T_2$     & $D$-$T_1$-$T_2$ & $D$-$T_1$-$T_2$ & $D$-$T_1$-$T_2$ \\
	SNR			       & 60				   & 100             & 87              & 187             \\
	$\chi^2$       & 1.01				 & 1.23            & 1.3             & 1.0             \\
	 $t_{comp}$ (s)       & 400			   & 5684            & 5955            & 5184            \\
	$\alpha_0$     & 1 				   & 0.05            & 2               & 0.01            \\	
	$\alpha_2$     & 0 				   & 0               & 0.1             & 0               \\	
	$N_{T_1}$      & 150 			   & 100             & 150             & 100             \\	
	$N_{T_2}$      & 150 			   & 100             & 150             & 100             \\	
	$N_{D}$        & - 				   & 100             & 150             & 100             \\	
	$N_{echo,bin}$ & 46296 			 & 302             & 103             & 300             \\	
	$N_{T_W}$      & 30	         & 14              & 12              & 13              \\	
	$N_g$	         & -           & 31              & 11              & 8               \\	
	$m_{LH}$       & 250 GB      & 1049 GB         & 367 GB          & 250 GB          \\	
	$m_{Bil}$      & 66.9 MB     & 9.3 MB          & 27.3 MB         & 8.5 MB          \\	
	$m_{LH}/m_{Bil}$  & $3.738\times10^3$ & $1.124\times10^5$ & $1.347\times10^4$ & $2.934\times10^4$ \\	
\hline 
 \end{tabular}
\caption[6 Multilinear Inversion Summary]{Inversion parameters and results. SNR denotes the signal-to-noise ratio; $\chi^2$ is the normalized sum of the square of the difference between the experimental data and the one predicted by the inversion; $t_{comp}$ the total time of computation; $\alpha_0$ and $\alpha_2$ are the zeroth- and second-order regularization parameters; $N_{T_1}$, $N_{T_2}$, and $N_D$ are the number of longitudinal and transverse relaxation rate points, and the number of diffusion coefficient points, respectively; $N_{echo,bin}$ is the number of binned echoes used for inversion; $N_{T_W}$ and $N_g$ are the number of polarization steps and magnetic field gradient steps, respectively; $m_{LH}$ and $m_{Bil}$ denote the memory required by the Lawson-Hanson and the bilinear inversion methods, respectively; and $m_{LH}/m_{Bil}$ is the ratio between the memory required by the LH and bilinear methods.}
\label{tab:Summary}
\end{table}

%% file: Conclusions.tex
\section{Conclusions}
\label{sec:Conclusions}

We introduced a new method to perform memory-efficient multidimensional NMR inversion. By generalizing the concept of a gradient from a vector to a tensor gradient, we avoid memory-expensive Kronecker products of dense matrices and perform multidimensional NMR multilinear inversion without kernel compression. Our work shows that: 

\begin{itemize}
\item Multilinear inversion yields results that are comparable to the Butler-Reeds-Dawson and Lawson-Hanson methods. 
\item Multilinear inversion is fast: the speed of steepest-descent nonnegative multilinear inversion is comparable to or faster than the Lawson-Hanson method.
\item Multilinear inversion is memory-efficient: the memory requirements for multilinear inversion is several orders of magnitude lower than that of the Lawson-Hanson and Butler-Reeds-Dawson methods.
\item Multilinear inversion is flexible: it can handle non-separable kernels, linear constraints, arbitrary regularization terms, and can be easily adapted to higher dimensions.
\item Multilinear inversion is easy to implement: only the cost function and its first derivative are needed in the inversion algorithm.
\end{itemize}

As illustrated in Figure \ref{fig:Figure_MethodsComparison}, to use Kronecker products, it becomes necessary to implement kernel compression via SVD to overcome the high memory requirements. With multilinear inversion, one has the option to use or not to use SVD compression, as it does not require Kronecker products and is therefore memory efficient. Multilinear inversion can be further sped up by using SVD compression, where possible.

%% file: AbbrevSymb.tex
\section{Acronyms and Symbols}
\label{sec:AbbrevSymbols}

\begin{table}[htb]
\footnotesize
\centering
 \begin{tabular}{c l} 
 \hline\hline
 Symbol & Meaning \\ [0.5ex] 
 \hline
$\tensor{K_1}$, $\tensor{K_2}$ & NMR kernels  \\
$\tensor{F}$, $\tensor{D}$ & distribution matrix, data matrix \\
$\tensor{K_A}$, $\tensor{F_A}$, $\tensor{D_A}$ & augmented kernel, augmented distribution, augmented data \\
$\nabla(C(\tensor{F_A}))$ & one-dimensional augmented gradient\\
$\nabla(C(\tensor{F}))$ & multi-dimensional tensor gradient\\
$\Re^{(M \times N)}$ & space of $M \times N$ real matrices \\
$||\cdot||_F$ & Frobenius norm \\
$m_A$, $m_{LH}$ & memory required by the augmented matrix, Lawson-Hanson \\
$m_{TG}$, $m_{Bil}$ & memory required by the tensor gradient, bilinear method\\
$N_A$ & number of operations required to calculate \\
 & the augmented gradient \\
$N_{TG}$ & number of operations required to calculate \\
 & the tensor gradient\\
$T_1$, $\tau_1$ & longitudinal relaxation rate \\
$T_2$, $\tau_2$ & transverse relaxation rate \\
$D$   & diffusion coefficient \\
$\alpha_0$, $\alpha_2$ & Tikhonov regularization coefficients: zeroth and second order\\
$\tensor{f}_\ell$ &  current solution during minimization\\
$\gamma_\ell$ &  current minimization step-size\\
$\Gamma\{\gamma_\ell\}$ &  set of minimization step-sizes\\
$\tensor{d}_\ell$ &  current direction during minimization\\
$\tensor{e}_\ell$ &  current error during minimization\\
$\tensor{P}$ &  projection operator \\
$\mathcal{L}$ & line-search scalar function \\ 
$N_{T_1}$ & number of $T_1$ points \\
$N_{T_2}$ & number of $T_2$ points \\
$N_D$ & number of $D$ points \\
$N_{echo}$ & number of echoes in raw data\\
$N_{echo,bin}$ & number of echoes after binning raw data\\
$N_{T_W}$ & number of polarization steps \\
$N_g$ & number of magnetic field gradient steps \\
$\sigma^2$ & variance of raw data noise \\
$\chi^2$ & chi-squared \\
$\gamma$ & proton gyromagnetic ratio = $2.675\times10^8$ T$^{-1}$s$^{-1}$ \\
$g$ & magnetic field gradient pulse amplitude \\
$\delta$ & magnetic field gradient pulse duration \\ 
$\Delta$ & time between magnetic field gradient pulses \\
$T_E$ & time between successive echoes \\
\hline\hline
 \end{tabular}
\caption*{List of symbols}
\label{tab:Symbols}
\end{table}

\begin{table}[htb]
\footnotesize
\centering
 \begin{tabular}{c l} 
 \hline\hline
 Acronym & Full Meaning \\ [0.5ex] 
 \hline
 NMR & Nuclear Magnetic Resonance \\
 IR & Inversion Recovery \\
 PFGSE & Pulse Field Gradient Spin Echo\\
 CPMG & Car-Purcell-Meiboom-Gill \\
 SVD  & Singular Value Decomposition \\
 LH  & Lawson-Hanson \\
 BRD & Butler-Reeds-Dawson \\
 1D, 2D, 3D, N-D & one-, two-, three-, N-dimensional \\
 SNR & Signal-to-Noise Ratio \\
 MB  & Megabyte = $10^6$ bytes \\
 GB  & Gigabyte = $10^9$ bytes \\
 TB  & Terabyte = $10^{12}$ bytes \\
 TG  & Tensor Gradient \\ 
 Bil & Bilinear Inversion Method \\ \hline \hline
 \end{tabular}
\caption*{List of acronyms}
\label{tab:Abbreviations}
\end{table}